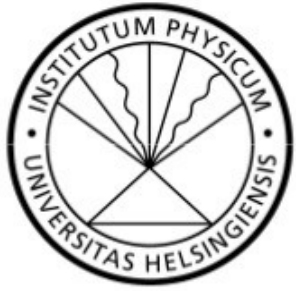

# Cosmic Topology

Jaspreet Sandhu
November 17, 2013

Tiivistelmä — Referat — Abstract

This thesis aims to cover the central aspects of the current research and advancements in cosmic topology from a topological and observational perspective. Beginning with an overview of the basic concepts of cosmology, it is observed that though a determinant of local curvature, Einstein's equations of relativity do not constrain the global properties of space-time.

The topological requirements of a universal space time manifold are discussed, including requirements of space-time orientability and causality. The basic topological concepts used in classification of spaces, i.e. the concept of the Fundamental Domain and Universal covering spaces are discussed briefly. The manifold properties and symmetry groups for three dimensional manifolds of constant curvature for negative, positive and zero curvature manifolds are laid out.

Multi-connectedness is explored as a possible explanation for the detected anomalies in the quadrupole and octopole regions of the power spectrum, pointing at a possible compactness along one or more directions in space. The statistical significance of the evidence, however, is also scrutinized and I discuss briefly the bayesian and frequentist interpretation of the posterior probabilities of observing the anomalies in a $\Lambda CDM$ universe.

Some of the major topologies that have been proposed and investigated as possible candidates of a universal manifold are the Poincare Dodecahedron and Bianchi Universes, which are studied in detail. Lastly, the methods that have been proposed for detecting a multi connected signature are discussed. These include ingenious observational methods like the circles in the sky method, cosmic crystallography and theoretical methods which have the additional advantage of being free from measurement errors and use the posterior likelihoods of models. As of the recent Planck mission, no pressing evidence of a multi connected topology has been detected.

# Acknowledgements

I would like to dedicate this thesis to my advisor, Prof Kari Enqvist who was always an inexhaustible source of motivation and eclectic advice. His enthusiasm took me a long way in keeping sight of the bigger picture while wading through a truly humungous pile of research literature. I would also like to thank Lauri for always enthusing me with new energy when needed and introducing me to the concept of taking regular breaks, to which i had long been oblivious, which unsurprisingly had a very positive impact on the quality of the thesis. I'd like to thank Valerie for her valuable critique on the thesis and most importantly, my working methods and general way of life. I would also like to thank Sebastian, Miguel, Åsa and Nasim for making my stay in Finland memorable, my times with them will always be close to my heart.

# Contents

# List of Figures

# Chapter 1

# Introduction

*The great charm resulting from this consideration (a multi-connected universe) lies in recognition of the fact that the universe of these beings is finite and yet has no limits.*
- Albert Einstein, on a multi-connected universe [1; 2]

In the year 1916, Einstein hypothesized the general theory of relativity, which exposed gravity as a manifestation of the curvature of space-time and further relegated the extent of curvature to the distribution of the surrounding mass. However, although this theory made it possible to predict the local geometry of space, it did not put any constraints on its boundary conditions or the global geometry of space-time.

Even though we encounter the significance of the spatial curvature in cosmology texts relatively often, the possibility of a multi-connected topology is usually not discussed. However, that does not imply that the possibility does not exist. Einstein's first seminal paper in cosmology discussed the possibility of a multi-connected universe according to a Clifford-Klein space form [3]. Einstein and Wheeler favored a finite universe on the basis of Mach's principle [4]; others have said that an infinite universe is unaesthetic and wasteful [5], while quantum cosmologists say that small universes have small action and are thus more likely to be created [6].

On different distance scales, the geometric representation of space changes, for instance, for distances between $10^{-18}$ and $10^{11}$ metres, it can be approximated by ordinary three dimensional euclidean space, $E^3$, while on larger scales of up to about $10^{25}$ metres, space is better described by a continuous Riemannian manifold where the curvature varies in response to the surrounding mass content.

Topology is essentially a study of continuity of surfaces and manifolds. It is a property which is preserved through kneading, stretching or squeezing the manifold, actions which change the metric. However, cutting, tearing or making handles in the manifold would result in a different topology. A famous example would be of a doughnut and a coffee cup, which are equivalent entities according to a topologist, since they can be continuously transformed into each other. These kind of transformations are called homeomorphisms and are discussed in detail in Chapter 4, along with other topological entities useful for classifying the universal manifold.

The first set of constraints that can be applied to limit the large number of possible topologies are homogeneity and isotropy, as is confirmed through observations which point at statistical isotropy and homogeneity in the distribution of astronomical objects [7]. However, a special class of models also exist that suspend the assumption of isotropy, Bianchi universes, which I discuss in Chapter 8. A better understanding of the best possible spaces that model our universe can be obtained by studying the vast range of abstract spaces available to us from topology. Cosmic Topology thereby combines these two areas, topology and cosmological observations, to form a better picture of what the universe looks like on scales comparable to its size.

The most promising source of cosmological data for the purpose of topology at the moment is the Cosmic Microwave Background, an ancient relic of the Big Bang [8]. With the advent of improved technology, we have been able to send satellites specifically for the purpose of measuring the microwave background, providing us crucial information on the content and structure of the universe about 300,000 years post big bang. This radiation contains tiny temperature fluctuations that reveal the pattern of density fluctuations in photon gluon plasma, before the photons decoupled and went on to stream freely through space, forming the CMB. These fluctuations were first detected by the Cosmic Background Explorer (COBE) in 1992 [9], for which George Smoot and John Mather received the 2006 Nobel Prize in Physics. High precision measurements were further made with the launch of another probe, the Wilkinson Anisotropy Probe (WMAP) in 2001 [10], which was 45 times more sensitive and had a 33 times higher resolution than the COBE satellite. The measurements were recently further improved with the Planck satellite [11], which was 10 times more sensitive and had 3 times the resolution of WMAP and mapped the sky in 9 instead of the previously used 5 frequencies.

The results obtained from the probes largely confirmed the predictions from the ΛCDM model (flat universe model) of the universe except for some anomalies in the large scale measurements. The quadrupole and the octopole were measured to be much lesser than the value predicted by the model and this has been linked to a possible compactification of dimensions in one or two directions in space [39]. This can be understood in analogy to a tied string which allows only a certain number of wave-modes to exist with a limit on the highest mode being the length of the string. As different harmonics come together to form a musical note, in a similar way the CMB is a combination of spherical harmonics of temperature fluctuations. The modes that are sustainable in an instrument give a lot of information about the shape of the instrument. For instance, weak second and fourth harmonics are characteristic of a clarinet. Hence, for a continuously defined field in space, we can split it into a sum of harmonics. These harmonics are eigen-modes of the Laplace operator, and in principle, vibrational modes of space. Hence just as the density fluctuations in a clarinet are a byproduct and hence a characteristic of its size and shape, the pressure and density fluctuations in the primordial plasma were constrained with the size and shape of the universe. Though we do not see the complete 3D modes we see their intersection with the 2D horizon sphere, and it is not hard to get the 3D modes from their 2D manifestations.

Some of the suggested manifolds currently in focus, which incorporate some of the observed anomalies, are the Poincare Dodecahedron, the Picard horn and the Bianchi universes. We also discuss three dimensional manifolds of constant curvature and the observational consequences of Euclidean, Spherical or Hyperbolic manifolds. A central premise of the thesis is the classification and properties of different topological manifolds and their relevance to our universe.

# Chapter 2

# Standard Cosmology

Cosmology essentially focusses on the study of the origin, structure and evolution of the universe. A central assumption in cosmology is that the universe is isotropic and homogenous at large scales, hereby referred to as the 'the cosmological principle'. It is however difficult to observe at what scale the premise becomes absolutely true, since even at large scales we observe small deviations from homogeneity and isotropy which are known to be seeds from which the large scale structures later emerged. The seeds themselves are believed to have originated from initial quantum fluctuations which were greatly magnified during the era of inflation. However, it would be safe to assume that the universe is statistically homogeneous at scales larger than about 100 Mpc.

Relativistic cosmology gives us an idea of the local properties that a manifold must possess as deduced from the Einstein's field equations. However, these equations, being partial differential equations, do not say anything about the global properties of space-time. Hence, a given local metric element could correspond to a large variety of topologically distinct universe models. Below are described some core concepts of cosmology which will be later utilized in subsequent chapters. More detailed discussions can be looked up in Peacock's [12] and Kolb's [13] classic texts on cosmology.

## 2.1 The Friedmann-Lemaitre-Robertson-Walker Model

A homogeneous and isotropic universe is one that can be sliced into maximally symmetric 3 spaces of constant curvature and these symmetries constrain greatly the possible allowed solutions to the global metric. Friedmann, Robertson and Walker proposed a metric which gives us all possible solutions for such constant curvature universes. The solutions include big bang solutions, de Sitter solutions and also includes those requiring a cosmological constant.

The Friedmann-Lemaitre-Robertson-Walker metric is of the form:

$$ds^2 = g_{\mu\nu}dx^\mu dx^\mu = -dt^2 + a^2(\eta)[\frac{dr^2}{1-\kappa r^2} + r^2(d\theta^2 + sin^2\theta d\phi^2)], \qquad (2.1)$$

where a(t) is the scale factor and $\kappa = 0, -1, 1$ corresponds to a flat, positively curved or negatively curved universe respectively.

Converting to conformal coordinates where the manifold can be thought of as a constant curvature manifold and the dynamics is incorporated into the conformal scale factor a($\eta$) using the definition of conformal time $d\eta = dt/a(t)$. The expression now becomes

$$ds^2 = a^2(\eta)[-d\eta^2 + d\sigma^2], \qquad (2.2)$$

where the spatial part of the metric is given by

$$d\sigma^2 = d\chi^2 + f(\chi)(d\theta^2 + sin^2\theta d\phi^2), \qquad (2.3)$$

$f(\chi)$ is a function of curvature given by

$$f(\chi) = \begin{cases} \chi^2 & r = \chi & \text{flat,} \\ sinh^2\chi & r = sinh\chi & \text{hyperbolic,} \\ sin^2\chi & r = sin\chi & \text{spherical.} \end{cases} \qquad (2.4)$$

Hence the FLRW metric of space-time can be expressed as

$$g_{\mu\nu} = \begin{bmatrix} -1 & 0 & 0 & 0 \\ 0 & \frac{a^2}{1-\kappa r^2} & 0 & 0 \\ 0 & 0 & a^2 r^2 & 0 \\ 0 & 0 & 0 & a^2 r^2 sin^2\theta \end{bmatrix} \qquad (2.5)$$

Calculating the Einstein tensor $G_0^0$ from the metric gives

$$G^0{}_0 = -3\frac{\dot{a}^2}{a^2} - 3\frac{\kappa}{a^2}, \tag{2.6}$$

$$G^0{}_0 = -\left(2\frac{\ddot{a}}{a} + \frac{\dot{a}^2}{a^2} + \frac{\kappa}{a^2}\right)\delta_{ij}, \tag{2.7}$$

$$G_{0i} = 0. \tag{2.8}$$

Substituting this along with the energy momentum tensor $T^\mu{}_\nu$ into the Einstein's equations will give us a system of equations relating different cosmological parameters. The energy momentum tensor for a maximally symmetric space-time has a simple form given by

$$T^\mu{}_\nu = \begin{bmatrix} -\rho & 0 & 0 & 0 \\ 0 & p & 0 & 0 \\ 0 & 0 & p & 0 \\ 0 & 0 & 0 & p \end{bmatrix} \tag{2.9}$$

where $\rho$ denotes the energy density and p the pressure and these quantities only depend on the time in a homogeneous universe.

## 2.2 Einstein's General theory of Relativity

Einstein related the curvature of space-time to matter with the local Einstein equations:

$$G_{\mu\nu} = 8\pi G T_{\mu\nu}, \tag{2.10}$$

where $G_{\mu\nu}$ represents a function of the metric of spacetime and $T_{\mu\nu}$ is the energy momentum tensor representing the matter distribution in space. These equations quantify the influence of matter fields on local curvature, however they do not determine the large-scale topology of the universe.

The Einstein's equation $G_{\mu\nu} = 8\pi G_N T_{\mu\nu}$ is a non-linear system of ten partial differential equations. In the case of a Friedmann-Lemaitre-Robertson-Walker (FLRW) universe, it reduces to two ordinary differential equations which can be rearranged to give us the *Friedmann equations* below:

$$3\frac{\dot{a}^2}{a^2} + 3\frac{\kappa}{a^2} = 8\pi G_N \rho, \tag{2.11}$$

$$3\frac{\ddot{a}}{a} = -4\pi G_N(\rho + 3p). \tag{2.12}$$

From these equations, one can obtain the energy continuity equation given by

$$\dot{\rho} = -3(\rho + p)\frac{\dot{a}}{a}.$$ (2.13)

From Equation 2.11 we deduce that the global energy density is a function of curvature. The correspondence is made in terms of the density parameter, $\Omega = \rho/\rho_c$ where $\rho_c$ is the total density of a flat universe and can be obtained by taking $\kappa = 0$ in eqn. 2.11, which gives us $\rho_c = 3H^2/8\pi G$.

Writing 2.11 in terms of $\Omega$, we get the expression $H^2 a^2(\Omega - 1) = \kappa$, where it becomes evident that $\Omega > 1$ for $\kappa = 1$ and $\Omega < 1$ for $\kappa = -1$. Hence the universe will be positively curved for $\kappa = 1$ and negatively curved for $\kappa = -1$ [14]

In terms of the curvature radius, $(R_{curv} = a/|\kappa|^{1/2})$, we have

$$R_{curv} = \frac{1}{H|\Omega - 1|^{1/2}}.$$ (2.14)

## 2.3 The Last Scattering Surface

About 300,000 years after the big-bang, the universe was a hot plasma of coupled baryons and photons. As the plasma cooled down to below the photon temperature, the photons decoupled from the mixture and started streaming freely through space thereby making it transparent. Thus, the surface of last scatter is the farthest that we can see into the universe, since that is the time when the photons decoupled from matter and were free to stream through space. So for a topology to be observable with current methods, a manifold's fundamental domain would have to be smaller than the diameter of the surface of last scatter (SLS).

It would thus be useful to calculate the radius of the surface of last scatter, which is the distance light has travelled from the time of decoupling to today,

$$D_\gamma = a(t) \int_{t_d}^{t_0} \frac{dt}{a(t)}$$ (2.15)

or in comoving units,

$$d_\gamma = \frac{D_\gamma}{a(t)} = \int_{t_d}^{t_0} \frac{dt}{a(t)} = \int_{\eta_d}^{\eta_0} d\eta = \Delta\eta.$$

The diameter of the SLS, $2\Delta\eta$ essentially represents the extent of the observable universe.

The volume of the observable universe can thus be determined by integrating the metric of space-time in radial coordinates,

$$V_{SLS} = \int \sqrt{-g} dr d\theta d\phi.$$

For $R_{SLS} = \Delta\eta$ in FLRW space-time metric as given in Equation 2.4, The corresponding volume of the observable universe is

$$\begin{aligned} V_{SLS} &= \pi(sinh(2\Delta\eta) - 2\Delta\eta), \quad \Omega_0 < 1, \\ &= \frac{4\pi}{3}\Delta\eta^3, \quad \Omega = 1, \\ &= \pi(2\Delta\eta - sin(2\Delta\eta)), \quad \Omega_0 > 1. \end{aligned} \tag{2.16}$$

For multi-connected spaces, the number of copies of the fundamental domain that can fit inside the observable universe would be given by [14] :

$$N = \frac{V_{SLS}}{V_M}.$$

where $V_{SLS}$ is the surface of last scatter and $V_M$ is the volume of the fundamental domain.

## 2.4 The Cosmic Microwave Background

The Cosmic microwave background though having being predicted since the 1940's was only accidentally discovered in 1964 by the American radio astronomers Arno Penzias and Robert Wilson who were later awarded the Nobel prize for the discovery. Probes were later sent to map the microwave background, namely COBE, WMAP and recently the Planck Satellite. The latest high resolution map of the CMB was release by the Planck satellite on 21st March, 2013 as shown in Figure 3.1 below.

The CMB is largely isotropic with tiny fluctuations which are now known to represent density perturbations in the early universe. The photons that produce the tiny fluctuations in the temperature spectrum are supposed to have originated at the surface of last scatter where they decoupled from baryons. The temperature fluctuations are made up of two components: the original temperature fluctuations in the baryon-photon fluid at the time of decoupling and the additional component introduced while on its journey to where we are today,

$$\left(\frac{\delta T}{T}\right)_{obs} = \left(\frac{\delta T}{T}\right)_{intr} + \left(\frac{\delta T}{T}\right)_{jour}. \tag{2.17}$$

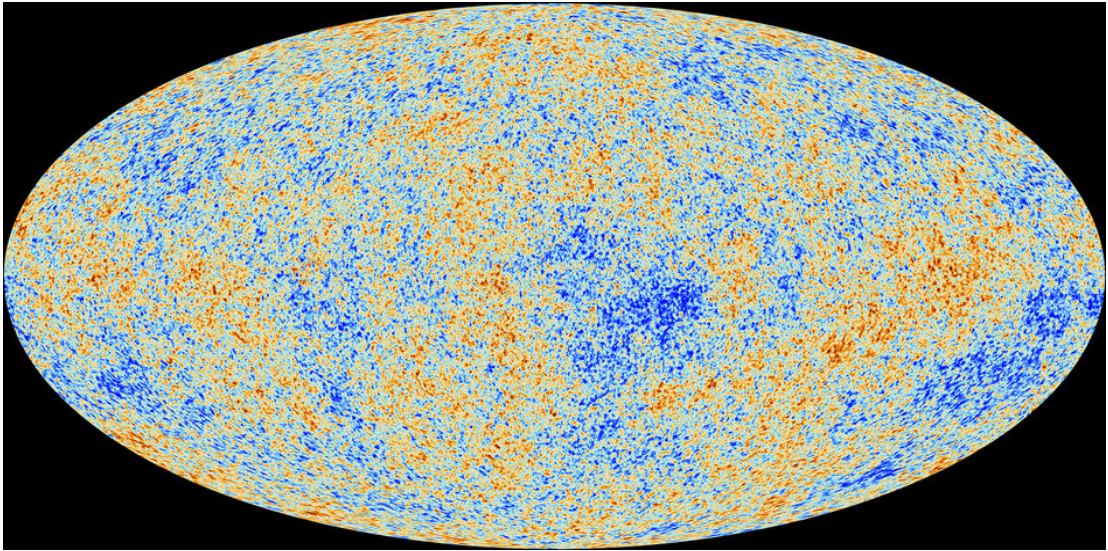

Figure 2.1: CMB temperature anisotropy map from Planck satellite.

The Temperature perturbations $\delta T$ can be defined either as the deviation from the averaged out temperature of the background or just the background temperature. The latter case is simpler and doesn't take the contribution of the fluctuations into the mean, though it is more common to use the former method.

The CMB radiation is uniform across the sky and exhibits small temperature fluctuations of 1 part in $10^5$. The temperature fluctuations can find their source in the frozen acoustic waves in the baryon photon plasma as the photons decoupled. The fluctuations in the photon temperatures arose from escaping from different density regions in the plasma, the photons from the denser regions lost more energy in escaping the plasma than the ones from rarer regions. The photons, in their journey towards us, became redshifted as well.

The temperature anisotropy can be expressed as a function on a sphere. In analogy to the fourier expansion in three-dimensional flat space, we can express the fluctuations as a expansion of spherical harmonics as

$$\frac{\delta T}{T_0}(\theta, \phi) = \sum_{l=1, m=-l}^{\infty, l} a_{lm} Y_{lm}(\theta, \phi). \tag{2.18}$$

The multipole coefficients, $a_{lm}$ can be calculated from:

$$a_{lm} = \int Y_{lm}^*(\theta, \phi) \frac{\delta T}{T}(\theta, \phi) d\Omega. \tag{2.19}$$

We take $a_{00} = 0$, since $Y_{00}$ is a constant and we are left with the average temperature. The $l = 1$ part represents the dipoles due to the motion of the solar system with respect to the last scattering surface interspersed with the actual cosmological dipole arising from large scale perturbations. Since it hasn't been possible to separate the two dipoles, the analysis of the CMB is taken from $l = 2$ onwards.

The multipole coefficients represent a deviation from the average temperature and are gaussian in nature having an expectation value of zero. The variance of the coefficients, $\langle |a_{lm}|^2 \rangle$ averaged over m gives us the average deviation. True to the isotropic nature of the background, we only observe dependence on l, which corresponds to the angular size of the mode and none on m, which represents orientation. Thus we have

$$C_l = \frac{1}{2l+1} \sum_m \langle |a_{lm}|^2 \rangle, \tag{2.20}$$

where $C_l$ is the angular power spectrum and contains all statistical information about the CMB. We compare the power spectrum predicted by different metric solutions to the one observed as a test of its validity.

The observed $C_l$ is the manifestation of a specific instance from an array of random possible outcomes. The observed angular power spectrum is the average of the observed $a_{lm}$ values,

$$\hat{C}_l \equiv \frac{1}{2l+1} \sum_m |a_{lm}|^2. \qquad (2.21)$$

On larger scales, the CMB begins to be dominated by the Sachs-Wolfe effect. It is given by

$$\left(\frac{\delta T}{T}\right)_{obs} = -\frac{2}{3}\Phi(t_{dec}, x_{ls}) + \Phi(t_{dec}, x_{ls}) + 2\int_{dec}^{0} \dot{\Phi}dt \qquad (2.22)$$

$$= \frac{1}{3}\Phi(t_{dec}, x_{ls}) + 2\int_{dec}^{0} \dot{\Phi}dt. \qquad (2.23)$$

The first term in the above equation, $\frac{1}{3}\Phi(t_{dec}, x_{ls})$ is the ordinary Sachs Wolfe effect and is thought to have originated due to a gravitational redshift at the surface of last scattering. The integrated Sachs-Wolfe effect on the other hand, $2\int_{dec}^{0} \dot{\Phi}dt$ originated between the surface of last scatter and the earth due to the presence of energy densities other than matter in the universe, for instance radiation energy or dark energy and thus is negligible in a universe dominated by matter density.

# Chapter 3

# Basic Conditions required of a universal manifold

We get a set of solutions from the general relativity equations which represent possible cosmological models. We can represent these models as an ordered pair $(M, \text{g})$ where $M$ is a connected smooth, n-dimensional manifold (n $\geq$ 2) and g is a smooth Lorentzian metric on M. The information on the topology of the universe is contained in M while g relates different points on the Manifold, M. The metric, g can be thought of as the function which assigns length to each vector $\xi^a$ in M, classifying them into three classes: timelike if $g_{ab}\xi^a\xi^b > 0$, light like if $g_{ab}\xi^a\xi^b = 0$ and space like if $g_{ab}\xi^a\xi^b < 0$ and gives them positive, zero and negative lengths respectively. The lightlike curves form a double cone structure at every point on the manifold, encompassing all timelike curves. The manifold, must admit a timelike vector field for a lorentzian metric to exist. A detailed analysis of causality conditions and further lower level constraints on a space-time manifold are discussed in [15; 16]

## 3.1 Time Orientable Manifolds

The world lines for a particle are represented with a cone like structure with two lobes for past and future oriented vectors. Special Relativity holds that a particle can only move on its worldline into the future and thus preserves the local orientability of time. This property holds also in curved spacetimes where general relativity takes over, with a straightforward extension of the local validity of special relativity. However, for a consistent global time orientation definition, local time orientations must be continuous and not clash, especially so, in closed trajectories, where time orientation must remain the same after a complete trip.

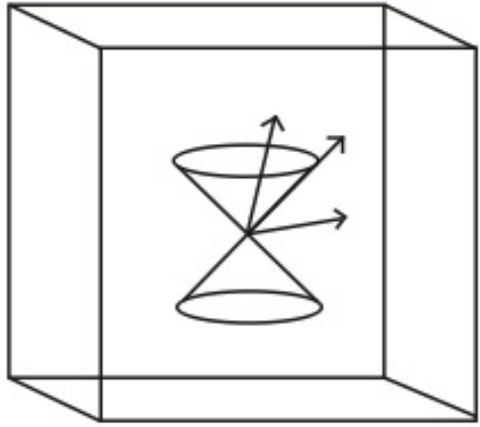

Figure 3.1: A double cone structure on a manifold with timelike, null and space like vector in, on and outside the cone respectively.

A simple test to check time orientability of a curve is to consider a point P lying on a curve C with an orientation along one of the direction, now take the point along the curve such that it returns to its previous position. If the orientation is still preserved, the curve is time orientable. A manifold is said to be time orientable, if every closed curve on the surface is time orientable.

## 3.2   Causal Manifolds

Notion of causality derives from the fact that cause precedes effect. We say that a point $p \in M$ chronologically precedes $q \in M$, *i.e.* $p \ll q$ if there is a timelike path from p to q. We can then define the future cone of p as

$$\mathcal{I}^+(p) = \{ \ q \in M : p \ll q\}. \tag{3.1}$$

For some interval, $I \subseteq \mathbb{R}$, a smooth curve $\gamma : I \to M$ is timelike if its tangent vector $\xi^a$ at each point in $\gamma$ is timelike. Similarly, a curve is null if its tangent vector at every point is null. A curve thus is causal if its tangent vector at each point is either null or timelike. For a space-time manifold to be physically possible, it should not have any closed timelike curves *i.e.* points which violate the chronology condition listed above. A manifold is further called *strongly causal* if for each p ∈ M there exist arbitrary small neighbourhoods U of p such that any causal curve that starts in U and then leaves U never re-enters U.

In special relativity, it implies that a real particle cannot travel along closed spacelike curves, which is another term for travel into the past. However in general relativity, it may be possible to have a view of the past while remaining in the future light cone, due to subtle distortions of spacetime by strong gravitational fields from a rotating massive body.

To find a stable solution, we can define stably causal manifolds. A spacetime is stably causal if it admits a cosmic time function that is a continuous real function and whose gradient is universally timelike. Some stably causal spacetimes are Minkowski, Schwartzchild, Friedmann, which are globally time orientable manifolds where time must increase along causal worldlines.

## 3.3  Global Hyperbolicity

A Cauchy surface S for M is an achronal (*i.e.* no timelike paths exist between any two points) hypersurface S in M which is met by every inextensible causal curve in M. Or simply put, it is the hypersurface constituting all of space at a given instant of time. Further, if S is a Cauchy surface for M, then M can be shown to be homeomorphic to $\mathbb{R} \times S$.
So in the simplest terms, Globally hyperbolicity can be defined as being equivalent to the existence of a Cauchy surface or the ability to foliate the manifold with Cauchy surfaces. Such a manifold is necessarily stably causal and time oriented.

The concept of determined path of evolution of a system from a certain point in space can be extended to spacetime through the concept of domain of dependence. For an initial spatial hypersurface $\Sigma$, the domain of dependence is a region of spacetime $D^+(\Sigma)$, such that any timelike curve reaching any point in $D^+(\Sigma)$ must intersect $\Sigma$. Similarly $D^-(\Sigma)$ is the past analog of $D^+(\Sigma)$. Hence we have a region of space-time

$$D(\Sigma) = D^+(\Sigma) \cup D^-(\Sigma),$$

which is entirely determined by information on $\Sigma$. Imposing this condition on spacetime infers that it is possible to determine the structure of spacetime from information on a single hypersurface, analogous to an initial value problem. It also implies that the interior of M can be taken to be diffeomorphic to $\mathbb{R} \times \mathbb{M}_3$ where $\mathbb{M}$ is a 3 dimensional Riemannian manifold.
A manifold without boundary is said to be globally hyperbolic if two conditions

hold:

1. For every pair of points p and q in the manifold, the space of all points that can be reached from p along a past oriented causal curve and from q along a future oriented causal curve, is compact and can be denoted by

$$D^-(p) \cap D^+(q).$$

2. Causality holds on the manifold.

## 3.4   Spatial Orientability

Additional restrictions could be imposed on spacetime based on space orientability. Space orientability can be easily described in terms of a two dimensional surface. On taking an arrow around the surface, if it returns to its initial position and still points in the same direction, the space is orientable and if not then it is non-orientable. A well known example of a non-orientable surface is the Mobius strip, which reverses the direction of the normal to the surface on complete trip around the strip, see Figure 3.2.

Generalizing to higher dimensional manifolds, we can define two types of curves, left handed and right handed. In three dimensions, a person crossing a non-orientable manifold, would have their left and right sides switched. Thus in general terms, a manifold which preserves the handedness of a closed curve is spatially orientable.

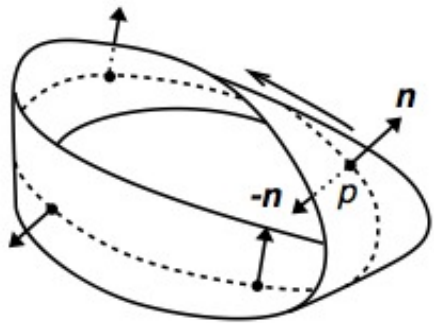

Figure 3.2: A Mobius strip is a good example of non-orientable 2-dimensional manifold. [15]

# Chapter 4

# Topological Classification of Spaces

Topology can be defined as the study of continuous transformations, albeit the property which remains unchanged when continuous transformations are made on a geometry, like squeezing, stretching etc. which change its metric but not its topology. Two manifolds that belong to the same topological class are called homeomorphic and can be continuously and reversibly transformed into each other. In other words, if we have two manifolds $M_1$ and $M_2$, a homeomorphism between them would be continuos map $\Phi : M_1 \mapsto M_2$ with a defined inverse.

FLRW Models admit spatial sections of homogeneous and isotropic spherical, hyperbolic or euclidean manifolds depending on whether the sign of spatial curvature is positive, negative or zero. However, there is often a common misconception that flat or hyperbolic universes imply an infinite universe, which was proved unfounded long ago by Friedmann [17] and Lemaitre [18] who discovered that FL metrics with zero or hyperbolic topologies did admit spatially closed topologies. However, the works of these and many other people remain ignored and cosmology textbooks implicitly assume space to be a simply connected hypersphere.

Given the isotropy of the microwave background, it is implied that the curvature of space is almost constant throughout. Hence literature on possible manifolds for the universe focuses mainly on manifolds of constant curvature. General relativity is invariant under diffeomorphisms which signify change of coordinates but not homeomorphisms. Thus, the principle of covariance goes a long way in predicting the laws of physics that a body follows in different regions of space, however topology remains independent of such a correlation.

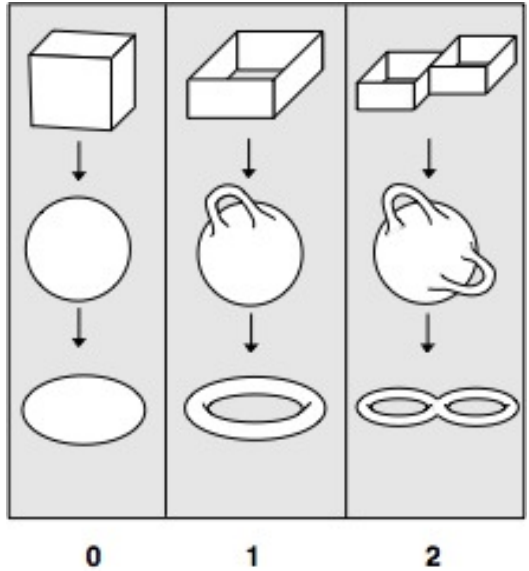

Figure 4.1: Some examples of homeomorphic manifolds, the numbers below the columns refer to the number of holes in the topology [15].

Below are explained some basic concepts that are frequently used in classifying and studying topological spaces:

## 4.1 Homogeneous Spaces

For two dimensional surfaces, it was shown by [19], that if a space is closed and connected, it is homeomorphic to Riemannian surfaces of constant curvature. A Riemann surface is defined as a complex manifold of dimension one. Hence all closed surfaces can be classified into one of the three Riemannian metrics: Spherical $S^2$, Euclidean $E^2$ and Hyperbolic $H^2$.

In three dimensions, this is not true, as we can clearly see from a three dimensional cylinder $S^2 \times R$ which is not homeomorphic to any of the constant curvature geometries, neither the spherical nor the euclidean manifold. The metric of the three dimensional cylinder is homogeneous but anisotropic. There are in total eight types of homogeneous three dimensional geometries out of which only five of them are of a constant curvature.

The symmetries of a manifold can be quantified with the group G of isometries, which are transformations to the manifold that leave the metric invariant. For a homogeneous manifold, G is non-trivial. The group H of all points y acts transitively on M as it relates to $g \in G$ such that $g(x) = y$ with $y$ being referred to as the orbit of $x$. The subgroup of isometries that leave the point $x$ fixed (for instance, rotations around x) is called the isotropy group, I at $x$. These isometries are related by

$$dim(G) = dim(H) + dim(I). \tag{4.1}$$

$G$ is simply transitive on $H$ if $dim(G) = dim(H)$ and multiply transitive if $dim(G) > dim(H)$.

The isometry group has dimensions $\leq n(n+1)/2$ for an n-dimensional manifold and attains maximum value for a maximally symmetric manifold. A maximally symmetric manifold is essentially a manifold which has the same number of symmetries as an ordinary euclidean space. For the space-time metric of our universe, a maximally symmetric space should have an isometry group of dimension = 10.

As [15] analyses, space times of dimensions ($\geq 6 \leq 10$) are not realistic cosmological models due the large number of dimensions involved. For dim(G) $\leq 6$, the group may act on $M$ or lower dimensional manifolds. For dim(G) = 4, the isometry acts on a manifold that is homogeneous in space and time and gives us a model which is spatially symmetric in space and time, but does not allow expansion. For the alternative case, when a subgroup of G acts only on the space-like hypersurfaces, giving us spatially homogeneous space times. This scenario has three possible sub cases:

- dim(G) = 6, where $G_3$ acts on spatially homogeneous spaces and there is a $G_3$ isotropy group. Thus, we have spatially homogeneous and isotropic space times which also allow space like hypersurfaces of constant curvature *i.e.* the FLRW models.

- dim(G) = 3, where there is only one group of isometries, *i.e.* the group $G_3$ acting on the spatially homogeneous spaces. Thus we have homogeneous and anisotropic spaces, one example of which are the Bianchi models [20].

- dim(G) = 4, where G is multiply transitive on 3 dimensional subspaces, some such space times are discussed in [21]. We will not consider these space-times in this text.

## 4.2 Simply and Multiply Connected Spaces

A good place to begin is the definition of the concept of homotopy which is an important classification regime used in topology. Two loops $\gamma$ and $\gamma'$ drawn on a manifold surface are said to be homotopic if one can be continuously transformed to the other. A simply connected manifold now can simply be defined as a manifold for which any loop is homotopic to another, or equivalently, all loops are homotopic to a point. If this condition is not true for all possible loops on the manifold, it is multi-connected. Homotopic loops give us information about holes or handles in a manifold. In higher dimensions, one dimensional homotopy loops are not enough to encompass all the properties of the topology, leading to the introduction of homotopy groups. The first homotopy group is called the fundamental group. Poincare conjectured that any connected closed n-dimensional manifold with a trivial fundamental group is topologically equivalent to a sphere [22].

Multiconnectedness implies that the fundamental group is non-trivial, essentially meaning that there is one hole in the manifold. Poincaré in 1904, conjectured that a connected closed n-dimensional manifold with a trivial fundamental group must be topologically equivalent to a sphere, $S^n$.

The set of solutions to Einstein's equations does not place any topological constraints on the manifold except its curvature. The FLRW models describe the observed universe with the greatest accuracy among the known models and give solutions for homogeneous and isotropic models with spherical, hyperbolic or flat topologies further incorporating a wide variety of possible solutions like the de Sitter solution or solutions with a cosmological constant or a non standard equation of state. The assumption in most literature of a simply connected universe is arbitrary and replacing the same with a multi-connected universe changes a very few characteristics in the FLRW models. One of the differences lies in the range of the coordinates where for a simply connected universe, one would have $\phi : 0 \to 2\pi, \theta : -\pi/2 \to \pi/2, \ \chi : 0 \to \infty$ for $k = -1, 0$ and $0 \to \pi$ for $k = 1$ whereas for a multi-connected model, space is smaller and the range of the coordinates is reduced.

As discussed earlier, observations suggest that the universe is homogeneous and locally isotropic which implies that space has constant curvature. Thereby, most multi-connected models explored in literature rely on this assumption with the exception of Bianchi and Lemaitre-Bondi models among a few others. However, even with the anisotropic models, the homogeneity and local isotropy of these models ensure that the CMB remains isotropic. A significant difference however is observed in the spectrum of density fluctuations.

While the finiteness of simply connected models can simply be determined from the sign of curvature of the manifold, *i.e.* infinite for $k = 0, -1$ and finite for $k = 1$, the same does not hold true for multi-connected topologies. As early as 1924, it was known that multi-connected models with a zero or negative curvature admitted spatially closed topologies [23; 24]. For instance a toroidal universe is of a finite volume and circumference despite zero curvature.

## 4.3    Fundamental Domain

A simple example is a torus whose fundamental domain is a rectangle. To obtain a torus from a rectangle, we first identify one pair of opposite sides in the rectangle, thereby getting a cylindrical tube. Identifying the other pair of opposite sides gives us a torus.

The transformations done in identifying the opposite edges form a holonomy group. A holonomy group is a subset of the full isometry group of the covering space. To understand the holonomy group, consider a point $x$ and a loop $\gamma$ at $x$ in the manifold $M$. If $\gamma$ lies in a simply connected domain of $M$, it generates a single point $\tilde{x}$ in $\tilde{M}$ but if the manifold is multi-connected, it creates a set of points $\tilde{x}'$, $\tilde{x}''$... which are said to be homologous to $\tilde{x}$. The displacements form isometries referred to as the holonomy group $\Gamma$ in $\tilde{M}$. Since there is a non zero distance between the homologous points, the group is discontinuous and has no fixed generating point. Thus, the holonomy group is said to act freely and discontinuously on $\tilde{M}$ [15]. An example of homologous points on a 2 torus manifold is shown in Figure 4.2

The full isometries of the universal cover are broken by identifications and we can represent a compact manifold as a quotient space given by G/$\Gamma$ where G is the group of isometries of the domain and $\Gamma$ is the holonomy group.

## 4.4    Universal Covering Space

The universal cover of a connected topological space is a simply connected space with a map f: Y → X that is a covering map. By acting with the transformation group on the fundamental domain, we get many identical copies of the domain which give us the universal covering manifold. In case of simply connected spaces this universal covering is identical to the fundamental domain for instance a sphere $\mathbb{S}^2$ is its own universal cover, however in the case of multiply connected spaces we get replicas of the central manifold. In this case, a universal manifold is constructed as follows. The space is cut open to make a simply connected space with edges, called the fundamental domain of the manifold. For instance, a hyperbolic octagon for a double torus and a square for a square torus. Now add another copy of the fundamental domain to the edge and keep doing so until all edges of the original manifold are covered. More copies are added to the resulting space recursively until a covering map with possibly infinite number of copies of the fundamental domain is obtained. The largest such possible cover is called the universal covering space. Thus, If f : Y → X is a covering map, then there exists a covering map f: $\tilde{X}$→ Y such that the composition of X and $\tilde{X}$ is the projection from the universal cover to X [25].

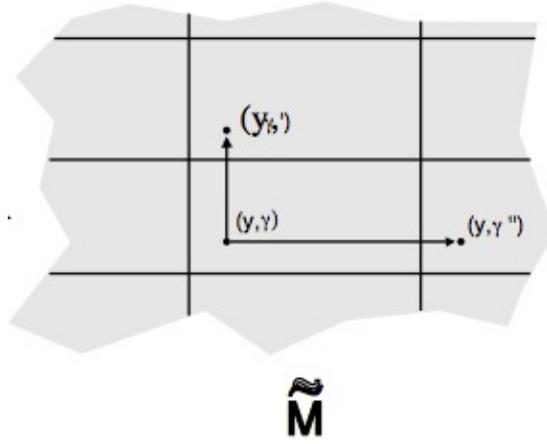

Figure 4.2: A representation of holonomic generators in a 2-torus manifold.

For example, a flat torus tiling a universal covering space can be likened to the screen of video game where on walking off the right end of the screen, one would emerge on the opposite edge and the same for the vertical edges. Thus, one gets the impression of an infinite space even though it is just a repetition of the same fundamental domain over and over. This type of a universal covering space is constructed by identifying the edges of a fundamental domain, and identifying the edges differently gives rise to a different set of orientations and symmetries on transition between different universes. An example of a universal covering of a 2-torus is given in Figure 4.4

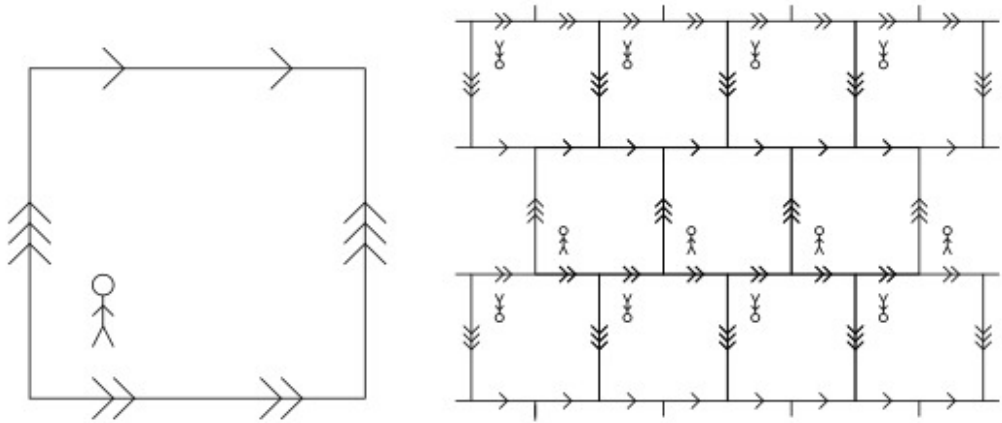

Figure 4.3: The universal covering space of a 2-torus, the space is tiled with squares which bear specific orientation and symmetry relations to each other [14].

## 4.5 Detectability of a Multi Connected Topology

There are basically three possible correlations between detectability of topology and the size of the universe. First, that the universe could be infinite in which case, it is not possible to detect topology with currently known methods. Secondly the universe is finite but much larger than the scale of the observable universe in which case, it is hard to detect visible signs of topology. The last and the best scenario would be a universe which is finite and comparable to the size of the observable universe, where we can thus use current methods discussed in Chapter 8 to detect its structure.

For a manifold M, there can be defined the smallest and largest circles inscribable in M, $r_{inj}$, described in terms of the smallest closed geodesic, $l_m$ and $r_{max}$ respectively. A closed geodesic, that passes through a point $x$, in a multiply connected manifold is a part of the geodesic that connect that point to its image in the covering space $\tilde{M}$. The length of any such closed geodesic which passes through $x$, in a manifold with a fixed isometry g, is given by the distance function:

$$\delta g(x) = d(x, gx). \tag{4.2}$$

In terms of this distance, the injectivity radius can be defined as

$$r_{inj}(x) = \frac{1}{2} min_{g \in \tilde{\Gamma}} \{ \ \delta g(x) \},$$ (4.3)

where $\tilde{\Gamma}$ denotes the covering group without the identity map.

We can define the observation survey depths to be $\chi_{obs}$. A topology is said to be undetectable if $\chi_{obs} < r_{inj}$ in which case we cannot detect any multiple images in the observable sky, and detectable for $\chi_{obs} > r_{inj}$.

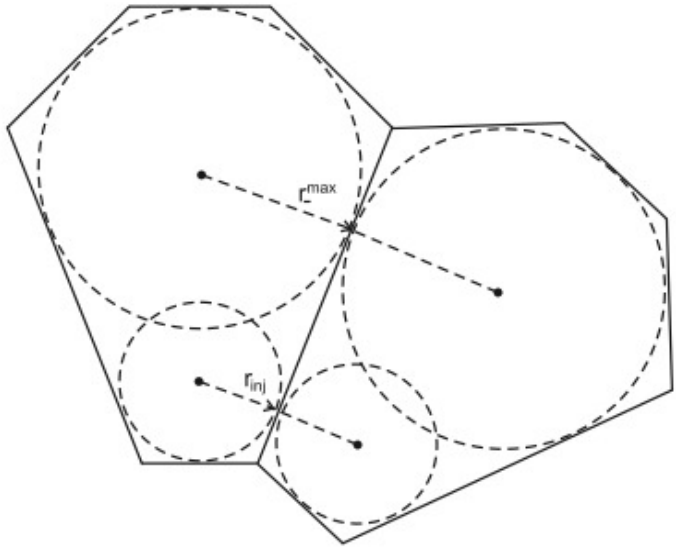

Figure 4.4: Two fundamental cells and the smallest and largest sphere's inscribable in M [26].

In a globally homogeneous manifold, the distance function for any covering isometry is constant, hence $r_{inj}$ is constant throughout space and the detectability of a topology does not depend on the observers position in the manifold, In inhomogeneous manifolds however, $r_{inj}$ varies from point to point and thus the topology depends on both the observers position and the survey depth. In this case however, we can still define an absolute undetectability condition, that is if for

$$r_{inj} = min_{x \in M} \{ \ r_{inj}(x) \},$$

$$\chi_{obs} < r_{inj}$$

then the topology is undetectable for all observers in M [26; 27].

For a flat manifold $E^3$, the relationship between the horizon radius and the injectivity radius is pretty arbitrary since it is possible to stretch its translational components to obtain any injectivity radius. It is thus probably least likely to be able to detect a Euclidean topology from the three possible topologies. On the other hand, in the case of a Hyperbolic topology, the volume of the fundamental domain increases as the complexity of the Hyperbolic group $\Gamma$ increases. However, the injectivity radius for the smallest hyperbolic manifolds exceeds the horizon radius. Spherical manifolds on the other hand decrease in size as the symmetry group $\Gamma$ becomes larger.

# Chapter 5

# Three Dimensional Manifolds of Constant Curvature

Current observations of the observable part of the universe imply a homogeneous and isotropic geometry to a precision of 1 part in $10^4$, thus we should begin by considering topologies that are locally homogenous and isotropic, thereby the constant curvature Euclidean, Spherical and Hyperbolic geometries as deduced in the previous chapter.

Any compact 3-manifold $\mathcal{M}$ with a constant curvature k allows a discrete isometry subgroup $\Gamma$ acting freely and discontinuously on $\tilde{\mathcal{M}}$. Such a manifold can also be written as $\tilde{\mathcal{M}}/\Gamma$, where $\tilde{\mathcal{M}}$ is the universal covering space of $\mathcal{M}$, given by Euclidean space (for $k = 0$), 3-sphere (for $k > 0$) or the hyperbolic 3-space (for $k < 0$). [15] does an excellent work of classifying the Euclidean, Spherical and Hyperbolic manifolds further into sub manifolds as discussed below.

## 5.1   Euclidean Manifolds

The line element of the Euclidean covering space is given by:

$$d\sigma^2 = R^2 \{\ d\chi^2 + \chi^2(d\theta^2 + sin^2\theta d\phi^2)\}. \tag{5.1}$$

where $\chi \geq 0$.

The full isometry group is given by $G = ISO(3) = \mathbb{R}^3 \times SO(3)$ and the generators of the possible holonomy groups $\Gamma$ include different combinations of identity, translations, glide reflections and helicoidal motions. In total 18 different types Euclidean manifolds can be generated. The manifolds can be classified primarily into open and closed models.

The open models include orientable and non-orientable space-forms which can be classified with glide reflections. On excluding glide reflection as a holonomy group, we get four orientable space-forms. The non-orientable space forms are not relevant to cosmology. The closed models on the other hand can be classified according to the different possible ways the opposite faces of a parallelepiped can be identified with each other. Another class of identifications can also be made on hexagonal fundamental polyhedron with rotations of $2\pi/3$ and $\pi/3$.

## 5.2   Spherical Manifolds

Spherical manifolds have a universal covering of a compact hypersphere. A 3-sphere S$^3$ of radius R is the set of all points in 4-Dimensional Euclidean Space. The metric of the 3-sphere with coordinates $x^0, x^1, x^2, x^3$ can written as

$$(x^0)^2 + (x^1)^2 + (x^2)^2 + (x^3)^2 = R^2. \tag{5.2}$$

Converting to angular coordinates $(\chi, \theta, \phi)$, for $\chi$ and $\theta : [0, \pi], \phi : [0, 2\pi]$

$$x^0 = R\cos\chi, \ \ x^1 = R\sin\chi\cos\theta, \ \ x^2 = R\sin\chi\sin\theta\cos\phi, \ \ x^3 = R\sin\chi\sin\theta\sin\phi.$$

We get the metric:

$$d\sigma^2 = R^2\{ \ d\chi^2 + \sin^2\chi(d\theta^2 + \sin^2\theta d\phi^2)\} \tag{5.3}$$

from

$$d\sigma^2 = (dx^0)^2 + (dx^1)^2 + (dx^2)^2 + (dx^3)^2.$$

The volume of the covering manifold is given by

$$vol(\mathbb{S}^3) = \int_0^\pi 4\pi R^2 \sin^2\chi R d\chi = 2\pi^2 R^3. \tag{5.4}$$

Substituting $r = \sin\chi$ in metric, we get the FLRW metric form of the spherical manifold

$$d\sigma^2 = R^2 \left\{ \ \frac{dr^2}{(1 - r^2)} + r^2(d\theta^2 + sin^2\theta d\phi^2) \right\}. \tag{5.5}$$

A good way of visualizing a 3-sphere is to use the analogy of a 2-sphere, where if we intersect the sphere with a plane and pass it through the sphere, the intersection points form circles of increasing diameter and subsequently of decreasing diameter. Similarly, one can imagine the intersection of a 3-sphere with a 3-dimensional hyperspace and forming spheres of increasing diameter before reducing in size again to zero.

The holonomy groups of $\mathbb{S}^3$ were classified by [28] into subgroups $\Gamma$ of SO(4) acting freely and discontinuously on $\mathbb{S}^3$:

- Cyclic group of order p, $Z_p(p \geq 2)$ : $Z_p$ can be seen as generated by the rotations by an angle $2\pi/\text{p}$ about some axis $[\theta, \phi]$ of $\mathbb{R}^3$

- Dihedral group of order 2m, $D_m(m \geq 2)$: Generated by rotations in the plane by an angle $2\pi/\text{m}$ and a reflection about a line through the origin. The operation preserves regular m-gons lying in the plane and centered on the origin.

-Polyhedral Groups: Symmetry groups of the regular polyhedra in $\mathbb{R}^3$ namely the Tetrahedral group of order 12, octahedral group of order 24 and Icosahedral group of order 60. The cube is included in the symmetry group of the octahedron and the dodecahedron is included in the symmetry group of the icosahedron.

There is an infinite number of spaces that can be obtained by taking the quotent of $\mathbb{S}^3$ with the above groups and varying the parameters p and m. The volume of the quotient manifold, $\mathcal{M} = \mathbb{S}^3/\Gamma$ obtained is given by

$$Vol(\mathcal{M}) = 2\pi^2 R^3/|\Gamma|. \tag{5.6}$$

## 5.3 Hyperbolic Manifolds

Some of the most important contributions to locally hyperbolic spaces were made by Thurston [29], however these manifolds still remain much less understood than other homogeneous manifolds. Nevertheless, $\mathbb{H}^3$ can be embedded in Minkowski space, $\mathbb{R}^3$ of metric

$$ds^2 = -(dx^1)^2 + (dx^2)^2 + (dx^3)^2 + (dx^4)^2$$

as the hypersurface,

$$-(x^1)^2 + (x^2)^2 + (x^3)^2 + (x^4)^2.$$

Thereby, the generators of the fundamental group G of $\mathbb{H}^3$ can be related to homogeneous Lorentz transformations [30].

Let us make coordinate transformations to introduce $(\chi, \theta, \phi)$ with $\chi \in [0, \inf), \theta \in [0, \pi], \phi \in [0, 2\pi]$, we get

$$x^1 = R \cosh \chi, \ \ x^2 = R \sinh \chi \cos \theta, \ \ x^3 = R \sinh \chi \sin \theta \cos \phi, \ \ x^4 = R \sinh \chi \sin \theta \sin \phi.$$

We thereby get the metric for $\mathbb{H}^3$ as:

$$d\sigma^2 = R^2 \left\{ \ d\chi^2 + \sinh^2 \chi (d\theta^2 + \sin^2 \theta d\phi^2) \right\}. \tag{5.7}$$

This metric can be expressed in a more commonly used, FLRW form, of the metric, obtained by the coordinate change, $r = \sinh \chi$ that gives us:

$$d\sigma^2 = R^2 \left\{ \ \frac{dr^2}{1 + r^2} + dr^2(d\theta^2 + \sin^2 \theta d\phi^2) \right\}. \tag{5.8}$$

The holonomies of $\mathbb{H}^3$ can be described as the group of fractional linear transformations acting of the complex plane:

$$z' = \frac{az + b}{cz + d}, a, b, c, d \in \mathbb{C}, ad - bc = 1.$$

This group is equivalent to the group of conformal transformations of $\mathbb{R}^3$ which leave the upper half space invariant.

The hyperbolic geometries for dimension $> 2$ are different from the 2-dimensional case. For instance, while a surface ($genus \geq 2$) can support an infinite number of non-equivalent hyperbolic metrics, a connected oriented manifold ($n \geq 3$) can only support at most one hyperbolic metric. This is confirmed further by the rigidity theorem which confirms that if two hyperbolic manifolds (dimension, $n \geq 3$) have isomorphic fundamental groups, they are necessarily isometric to each other. Hence for $n \geq 3$, the volume of a manifold and the lengths of its closed geodesics are topological invariants.

For compact euclidean spaces, the fundamental polyhedron can possess only a maximum number of eight faces, despite allowance of arbitrary volume. In spherical manifolds, the volume needs to be finite and a fraction of the maximum possible volume $\mathbb{S}$, *i.e.* $\mathbb{S}/\Gamma$. For the case of hyperbolic manifolds however, there is no limit on the possible number of faces of the fundamental polyhedron. There is a lower limit however, on the minimum volume of the hyperbolic 3-manifold, a lower limit of which was set by Meyerhoff [31] to be $Vol_{min} > 0.00082R^3$.

# Chapter 6

# Evidence in favor of Multi-connectedness

In 1900, Schwarzschild was already contemplating on whether the universe could have a finite and non-trivial topology. With the advent of modern technology, it has become feasible to test suggested theories and disregard or modify the ones that prove contradictory to observational evidence. The measurements of the temperature fluctuations in the CMB provided a solid observational basis to test current cosmological models. With the first set of data from the Planck satellite available, we have come a long way from COBE, which provided us with the first set of data quantifying perturbations in the microwave background.

The observations from WMAP and Planck confirm the predictions made by COBE [32] as well as confirmed the observed anomalies in comparison with the expected results if the ΛCDM model is deemed correct. The anomalies were especially evident on larger scales where the observed values of the multipoles were found to be much smaller than the theoretical values. This suppression of power has been widely explored in literature [34], [35], [36], [37], [38] as a possible evidence of the compactness of space. As pointed out in the introduction, a closed manifold would only allow certain harmonics to exist and the harmonics larger than the compact dimension will be suppressed, which would show in observations as a suppression in large scale power. Therefore, it is possible that the CMB anomalies observed at large scales could be a signal of the non trivial topology of the universe.

However the significance of these results as an invalidation of the concordance model is still under debate. It was shown by the WMAP team that the chances of the above anomaly occurring in a ΛCDM universe are one in 143 for the quadrupole and one in 666 for the overall large scale power. It has been shown by Efstathiou [33] that these anomalies do not significantly deviate from the concordance model employing bayesian and frequentist methods as will be discussed below in section 6.3. Also, Oliveira et. al. [39] in their paper point out that the galactic cut that was used to remove foreground contamination from the map may have included some of the strongest hot and cold spots of these multipoles.

However, a small universe is not the singular possible explanation for the anomalies. There are other models also which have ben put forward to explain the low large scale power, for instance a cutoff in the primordial power spectrum [41], possibly linked to the spatial curvature [40], multi field inflation models [42; 43] and quintessence models which involve a partial cancellation of the usual integrated Sachs-Wolfe effect [44].

## 6.1 Low Quadrupole

The surprisingly small CMB quadrupole was first observed by COBE [45] and later confirmed by WMAP settling the initial doubts of the results being possibly noise induced. The results were further confirmed with the application of a refined analysis of Galactic foreground contamination. However, there still remains some uncertainty about the contribution of the cut in making the anomaly seemingly more striking. The observed quadrupole is the sum of cosmic and dynamic quadrupole and the contribution of the dynamic quadrupole can be determined from the dipole. A generic quadrupole has three orthogonal pairs of extrema (two maxima, two minima and one saddle point). However, the quadrupole that has been observed has its strongest pairs of lobes very close to the galactic plane, which could have removed a large fraction of the quadrupole power spectrum with the application of the galactic cut. The saddle point is close to zero and the quadrupole has a preferred axis in space along which the quadrupole has no power.

The noise variance and beam issues are really low at large scales. Additionally the galactic foreground is only present in a small fraction of the sky and thus contributes negligibly to the total power spectrum. The WMAP team measured the quadrupole only from the part of the sky outside the galactic cut and hence the dominant uncertainty in its value was attributed to the foreground modeling. However, to get a better estimate of the effects of the galactic foreground, [46] measured the power spectra of the uncleaned ILC map and then subsequently zeroed out the power in the dirtiest regions successively leaving only the three cleanest regions. Band pass filtering of the resulting map produced almost the same spatial plots and power spectrum of the lowest poles as observed in the ILC map, signifying that galactic modeling does not play a large role in the uncertainty of the multipoles.

Using different combination maps and foreground cuts, we obtain slightly different values of the quadrupole though they are all very low when compared to the expected theoretical value according to the $\Lambda CDM$ model. A table comparing the theoretically expected and measured values obtained from independent methods is given below:

Table 6.1: Comparision between the model and the measured values of the quadrupole and octopole [46].

| Measurement | $\delta T_2^2 [\mu K^2]$ | p-value | $\delta T_3^2 [\mu K^2]$ |
|---|---|---|---|
| Spergel et al. model | 869.7 | | 855.6 |
| Hinshaw et al. cut sky | 123.4 | 1.8% | 611.8 |
| ILC map all sky | 195.1 | 4.8% | 1053.4 |
| Cleaned map all sky | 201.6 | 5.1% | 866.1 |
| Dynamic quadrupole | 3.6 | | |

As evident from Table 6.1, Tegmark et. al. [46] obtained largely the same value for the quadrupole than the ILC maps. Thus even though Tegmark uses completely different modeling techniques, the quadrupole turns out to be virtually identical to the WMAP teams quadrupole which strengthens the argument that the contribution of galactic foregrounds to the quadrupole power spectrum is not significant. The p-value gives the probability that the given value of quadrupole is obtained given that the $\Lambda$CDM model is true and is calculated from a $\chi^2$ distribution for $\delta T^2$ with 5 degrees of freedom.

## 6.2 Quadrupole-Octopole alignment and Planar Octopole

Both the quadrupole and the octopole appear quite planar with their hot and cold spots centered on a single plane. To quantify the significance of the alignment, in [39], the angular momentum dispersion about an axis $\hat{n}$ was defined by

$$\langle \psi | (\hat{n}.L)^2 | \psi \rangle = \sum_m m^2 |a_{lm}(\hat{n})|^2, \tag{6.1}$$

where $a_{lm}(\hat{n})$ refers to the spherical harmonics in a rotated coordinate system with its z axis in the $\hat{n}$ direction, see Table 6.2. The axis around which this dispersion is maximized was then calculated. The preferred axis for the quadrupole and the octopole come out to be

$$\hat{n}_2 = (-0.1145, 0.5265, 0.8424),$$

$$\hat{n}_3 = (-0.2578, -0.4207, 0.8698),$$

which is roughly in the direction of $(l, b) \sim (-110°, 60°)$ towards the Virgo cluster.

Table 6.2: Observed real values of quadrupole and octopole $a_{lm}$ coefficients in Galactic and Rotated(in direction of alignment) coodinates for the Cleaned map from [46].

| l | m | Galactic | Rotated |
|---|---|----------|---------|
| 2 | -2 | -21.43 | 13.32 |
| 2 | -1 | 6.03 | 0.40 |
| 2 | 0 | 10.73 | 6.72 |
| 2 | 1 | -8.30 | -0.40 |
| 2 | 2 | -19.39 | 28.86 |
| 3 | -3 | 40.71 | 50.58 |
| 3 | -2 | 2.45 | -1.67 |
| 3 | -1 | 0.96 | -0.50 |
| 3 | 0 | -6.52 | -13.60 |
| 3 | 1 | -12.84 | -0.27 |
| 3 | 2 | 30.50 | 0.71 |
| 3 | 2 | -19.29 | -20.68 |

If the CMB is assumed to be an isotropic field, $\hat{n}_2$ and $\hat{n}_3$ should be independent of each other and subsequently the dot product between $\hat{n}_2$ and $\hat{n}_3$ should be randomly distributed between [-1,1]. In reality, the observed dot product between them comes out to be approximately equal to 0.9838. Thus, they are inclined to each other at an angle of 10.3°. The probability of an alignment this good happening is estimated to be 1 in 62 [39].

The octopole in the CMB has also been observed to be unusually planar, infact to a greater degree than the quadrupole. All other higher multipoles exhibit usual behavior as predicted by the $\Lambda$CDM model for a gaussian isotropic field.

The chances of the multipole being planar is 1 in 20 [39]. Oliveira et.al. [39] also defined a test statistic t, that measured the ratio of power from the m = |3| coefficients to the total power as

$$t = \frac{max_{\hat{n}}|a_{-3,3}(\hat{n})|^2 + |a_{3,3}(\hat{n})|^2}{\sum_{m=-3}^{3}|a_{3,m}(\hat{n})|^2}. \tag{6.2}$$

This statistic for the rotated coordinates from Table 6.2 gives us a ratio of 94%. Monte Carlo simulations with independent values of $a_{lm}$ coefficients corresponding to an isotropic Gaussian random field, we get higher t values than that calculated above only 7% of the time. Thus the contribution from $|m| = 3$ is dominant in the octopole. In comparison to the octopole, the quadrupole is not as significant in its planarity.

## 6.3   Statistical Significance of the WMAP results

The significance of the low quadrupole and planar octopole are a topic of discussion at present. Many factors are being taken into account which could have exaggerated the signal. For example, the effects of foreground contamination or alternative reconstructions of the true CMB signal from the five frequency band maps measured by the CMB might have contributed somewhat in skewing the original value.

Inspite of the importance of paying attention to discrepancies observed in the observational test of an otherwise well confirmed model, we need to be sure that these discrepancies pass the significance tests of validity. Another question that may be asked is whether we get better overall fits to other modified models. These questions were explored by Efstathiou [33] in his paper where he applies the frequentist and bayesian analysis to arrive at a more realistic value for the chances of getting the same value, given that the ΛCDM Model holds true.

It was shown by Spergel [47] that the observed values of the quadrupole and the octopole would occur with a probability of 1 in 700 if the concordance model was assumed to be correct. According to the current observations, there is an almost complete lack of signal at scales larger than 60°. Keeping this in mind, Spergel's [47] method was to calculate the statistic

$$S = \int_{-1}^{\frac{1}{2}} [C(\theta)]^2 d(cos\theta), \qquad (6.3)$$

for a large number of simulated CMB backgrounds for a $\Lambda$CDM universe. It was found that the probability of obtaining a value of S lower than the measured value, was only $1.5 \times 10^{-3}$, thereby providing strong evidence that the observed values do not firmly support the $\Lambda$CDM universe. Efstathoiu [33], in his paper, used Monte Carlo Markov chains to calculate the posterior probability of the quadrupole given the CMB power spectrum and the covariance matrix and higher multipoles ($800 \leq l \leq 2000$) from CBI [50], ACBAR [49] and VSA [48]. He followed the MCMC analysis of [41] which used a six parameter $\Lambda$CDM model, namely a constant scalar spectral index $n_s$, spectral amplitude $A_s$, Hubble constant h, baryon density $\omega_b = \Omega_b h^2$ , CDM density $\omega_c = \Omega_c h^2$ and redshift of reionization $z_{eff}$ and evaluated the posterior distribution of the quadrupole for the given parameter set.

The histogram of the resulting quadrupole amplitudes is shown in Fig. 6.1. The peak corresponds to the value, $\Delta T_2^2 = 1250 \ \mu K^2$ with few values above 2000 and below 1000 $\mu K^2$. This is quite a narrow distribution and relatively far form the observed value of $\Delta T_2^2 = 123 \mu K^2$ in the publicly available data release.

For a frequentist analysis, we take the distribution of $C_l$ which, for gaussian amplitudes of $a_{lm}$, will be given by a $\chi^2$ distribution, with further dependence on other factors such as the Galactic cut, instrumental noise and other sources of error

$$dP(C_l) = \left(\frac{C_l}{C_l^T}\right)^{\frac{2l-1}{2}} exp\left(-\frac{(2l+1)C_l}{2C_l^T}\right)\frac{dC_l}{C_l^T}. \qquad (6.4)$$

$C_l^T$ being the expectation value of $C_l$. Integrating Equation 6.4 gives us the probability of observing a value $\leq C_l$ as

$$P(\leq C_l) = \frac{\gamma\left(\frac{2l+1}{2}, \frac{2l+1}{2}\frac{C_l}{C_l^T}\right)}{\Gamma\left(\frac{2l+1}{2}\right)}. \qquad (6.5)$$

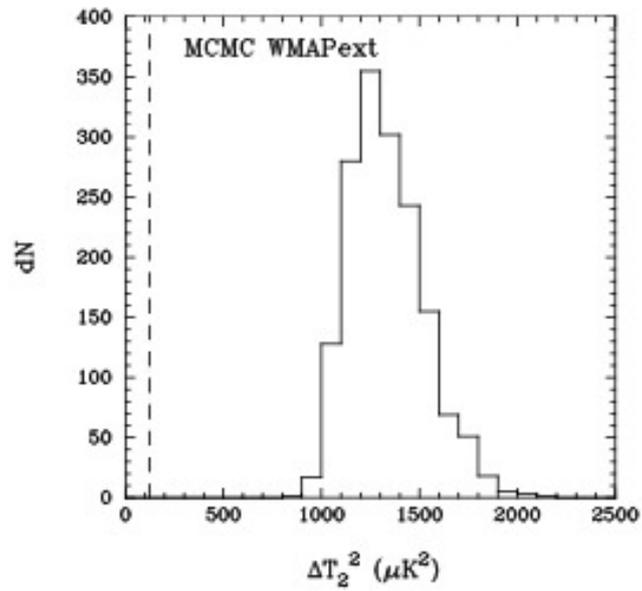

Figure 6.1: A histogram of quadrupole values obtained for MCMC simulation universes for a ΛCDM model; the dashed line is the value of the quadrupole observed by WMAP [33].

Efstathiou [33] applied the pseudo $C_l$ estimator to about $10^5$ simulations with a galactic cut of $\pm 10°$ and obtained a histogram of the estimated quadrupole values. Calculating the cumulative probabilities, the joint probability that $\Delta T_2^2 < 123 \mu K^2$ and $\Delta T_3^2 < 611 \mu K^2$ was determined to be 0.32% which is twice the probability calculated by Spergel et. al. [47]. With other small changes such as comparing the observed values with values from a lower allowed range of the theoretical spectrum or taking into account errors in the auto correlation function that come from the inaccurate subtraction of the galactic foreground should decrease the significance of the observations even further.

Efstathiou also applied bayesian analysis to the problem and calculated another set of significance level tests on the discrepancies between the observation and the model. A posterior probability analysis was done for the true values from the theoretical model manifesting a certain value on observation. According to the Bayes theorem, the posterior probability of a hypothesis H given data D is given by

$$P(H/D) \propto P(D/H)P(H), \qquad (6.6)$$

where P(H/D) is the probability of the hypothesis being true given a certain observation and P(H) is the prior probability of the hypothesis. Assuming the hypothesis that $C_l^T$ lies in the range $C_l^T \pm dC_l^T$, the posterior probability distribution for $C_l^T$ is given by

$$dP(C_l^T) \propto \frac{1}{\left(C_l^T\right)^{\frac{2l+1}{2}}} exp\left(-\frac{(2l+1)C_l}{2C_l^T}\right) dC_l^T, \qquad (6.7)$$

where $C_l$ is the observed amplitude and is also the maxima of the probability distribution curve of $C_l^T$.

Efstathiou [33] calculated the bayesian probability that the true values of the quadrupole and the octopole is greater that those predicted from the model, given the set of observations made ($\Delta T_2^2 = 123 \mu K^2$, $\Delta T_3^2 = 611 \mu K^2$ [51]). The discrepancies between the observed values and the peaks of the probability distributions of the true values further reduce in this analysis with the ratios being $p(C_2)/p(C_2^T)_{fid} = 28$ and $p(C_3)/p(C_3^T)_{fid} = 1.6$. Thereby, the frequency with which $\Delta T_2^2 > 1000 \mu K^2$, given the observed WMAP quadrupole of $123 \mu K^2$ is only 0.10, which translates to only 1 in 10 chances.

Thus though the visible discrepancies suggest new physics, they do not entirely rule out the $\Lambda$CDM model. Keeping this in mind, we move on to discuss the topological models suggested to better incorporate current observations and the methods being devised to enable their detection.

# Chapter 7

# Important Topologies

Spatial curvature gives us a first guess at the class of topologies that could be possible candidates to describe the universal manifold and can be directly calculated from an all sky analysis giving us a density parameter, $\Omega_{total}$ which measures the average mass energy density of space. As discussed in Chapter 2, depending on the whether the density parameter is less than, equal to or greater than 1, determines whether the topology will be negatively curved, flat or positively curved respectively. Before dark matter was discovered, the total density parameter $(\Omega_{tot})$ only amounted to around 0.3, leading to a widespread and almost exclusive analysis of negatively curved topologies as possible models for the universe. At present, the value of the density parameter is estimated to be $\Omega_{total} = 1.02 \pm 0.2$ at the $1\sigma$ level [52] leading to a situation where any of the three curvature topologies could be possible candidates, when taking observational errors into account.

The current estimate was made by the WMAP team, whose data turned out to be well within the predictions for the concordance model spectacularly except on the largest scales ($> 60°$)[52]. The observed quadrupole is close to 0.15% of the expected value, while the octopole was less strikingly low but still about 72% of the expected value. A quadratic maximum likelihood analysis gives, however, somewhat larger probabilities in the range 3.2-12.5 % [33]

If the universe were infinitely flat, the temperature fluctuations would have been expected to be present at all wavelengths just like a string of infinite length supports waves of all wavelengths. One of the arguments for the lack of such uniform presence of all temperature fluctuations is the compactness of space, thereby placing an upper limit on the wavelengths that can exist. Thus, cosmologists now face the problem of finding a model for space that while producing the said anomalies on large scales, also closely follows the near perfect flat space predictions on smaller scales.

We study below some of the more widely discussed topological models that could produce the spectra that we observed in the CMB.

## 7.1   Poincare Dodecahedron

With the current estimates of the density parameter tilting ever so slightly towards a value larger than 1, a positively curved model seems to draw more favor as a viable topology. The simplest alternative was considered first, that of a 3-torus, which was indeed found to suppress the quadrupole. However, it was also shown to lower values of other multipoles, which did not mirror observations [53]. Some other models studied corresponded to lens spaces $L(p, q)$, however these spaces suppressed the higher multiples even more than the lower multipoles, contradicting observations.

It was observed soon that only well proportioned spaces selectively suppressed the lower multipoles. The spaces that fall in this category are the binary polyhedral spaces $S^3/T^*, S^3/O^*, S^3/I^*$ which have positive curvatures and fundamental domains of a regular octahedron, truncated cube and a regular dodecahedron respectively. The Regular dodecahedron, commonly known as the Poincare dodecahedron was found to be within $R_{hor}/R_{curv} = 0.47$ for $\Omega = 1.02$. The feature that set apart this particular topology from the numerous other positively curved models, that haven't already been ruled out due to obvious discrepancies, is that for a certain range of the density parameter, the poincare dodecahedron exhibits a strong suppression of power at $l = 2$ and a weak suppression at $l = 3$ when normalized to the $l = 4$ power.

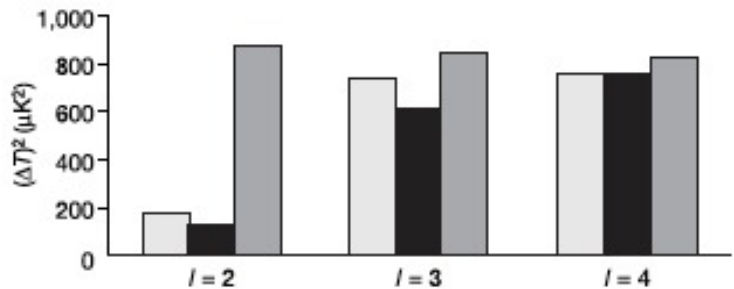

Figure 7.1: Comparision between multipoles from observation (light grey) and a flat (dark grey) and dodecahedral universe (black), from [54].

The Poincare dodecahedral power spectrum depends on the mass energy density parameter. It was found by Luminet et. al. [54] that the octopole power matches WMAP's best predictions if $\Omega_{tot}$ is between [1.010, 1.014], while for suppressed value of the quadrupole, $\Omega_{tot}$ lies in the range [1.012, 1.014] as shown in Figure 7.2. Thus, it is quite clear that WMAP's error range for the density parameter comfortably includes both of these intervals, [1.02 ± 0.2]. A comparison between the multipoles values obtained from the model to that from observation for $\Omega_m = 0.28$ and $\Omega_\Lambda = 0.734$ is shown in Figure 7.1.

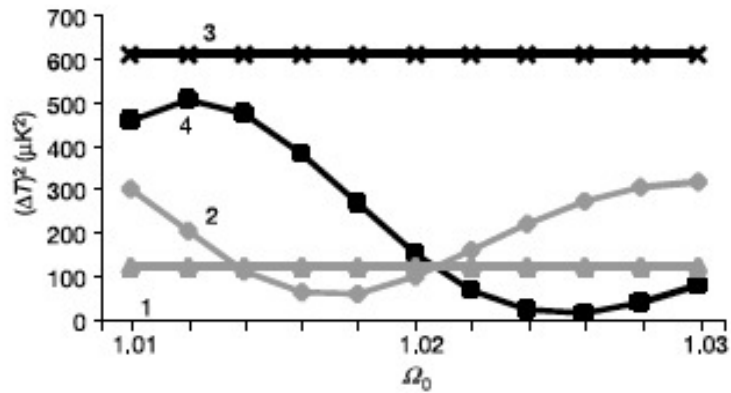

Figure 7.2: Range of density parameter, $\Omega_{total}$ for which the dodecahedral space agrees with WMAP observations, 1 and 3 being the observational data for quadrupole and octopole respective and 2 and 4 being the predictions from the dodecahedral model [54].

It is also striking that the Poincare dodecahedron does not have any degrees of freedom regarding its size or orientation. It is a positively curved space with a volume 120 times smaller than the simply connected hypersphere. It is fixed in its geometrical possibilities being a regular fixed volume dodecahedron and the only adjustable parameter is $\Omega_{tot}$. It is also globally homogeneous and thus looks the same to all observers at different positions. If we were to travel out of one side of the cell, we would return the same manifold albeit from the opposite face of the manifold. The adjoining edges (120°) of the spherical dodecahedron fit together snugly to tile the hypersphere.

Yet, however appealing the striking correlations with the observation may seem, strict observational tests are a necessity for its consideration as a serious possibility, the matched circles test being a viable test. Cornish et. al. [55] pre-dicted detectable temperature correlations in the CMB for small multi-connected spaces. The radius of the six matched circles increases with increasing $\Omega_{tot}$. If the density parameter were to be 1.013, our horizon radius being 0.38 in curvature radius units, then the in and out radii of the dodecahedron would be 0.31 and 0.39 respectively, and hence the volume of the fundamental cell only 83% of the observable sphere. The observable sphere in this case will intersect itself in six symmetrically placed circles of angular radius of about 35 degrees. However, no such circles were detected in the numerous independent searches made for the 7-year WMAP data [57].

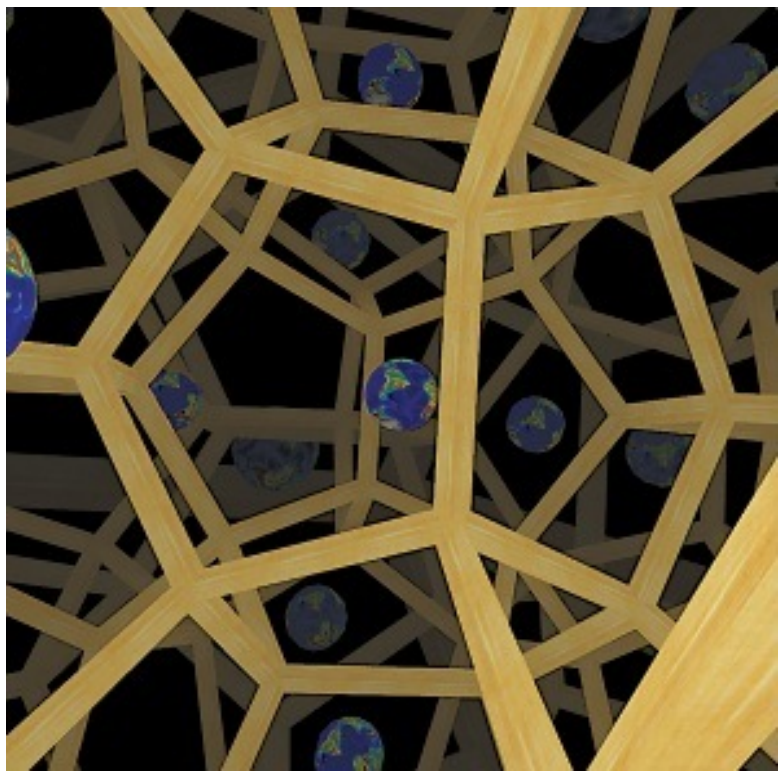

Figure 7.3: A Poincare Dodecahedron tiled universe [2].

## 7.2 Bianchi universes

One of the basic assumptions that the ΛCDM model relies on is that the universe is homogeneous and isotropic on large scales. However, if we relax the assumptions on the global isotropy of space-time, we would get more generalized solutions of the Einstein's field equations, giving us Bianchi universes. The re-cently released topological analysis from the Planck team discussed Bianchi $VII_h$ models in detail, given that no statistically significant pairs of circles were detected in the searches performed on the high resolution data[11].

If we consider small anisotropy, we can get our solutions by applying small perturbations to the standard FRW model to evaluate the subdominant contribution of the Bianchi component to the original temperature map from the $\Lambda$CDM Model [56].

The most general of the Bianchi Models are Bianchi $VII_h$ for flat, open models and Bianchi IX for closed models, and were analyzed in detail in [56] and [59] respectively. Bianchi $VII_h$ models were first compared to COBE in [62] and [63], and to WMAP in [60]. Jaffe et. al. [60] discovered a statistically significant correlation between one of the Bianchi $VII_h$ models and the WMAP ILC map by superimposing CMB from an unknown Bianchi component on the best-fit $\Lambda$CDM cosmology. Though promising, the parameters deduced from the Bianchi component did not agree with those from the $\Lambda$CDM. The important point to notice here though is that, on superimposing the Bianchi and $\Lambda$CDM data, some of the anomalies deduced in initial data disappeared.

Later, a variation of the above analysis, including dark matter was explored in [61]. The results, as before still did not provide consistency between the parameters derived from the Bianchi template and the CMB data. The first analysis of the scenario where the parameters of the Bianchi model are coupled to the parameters from the standard cosmological model was performed in [56].

Bianchi $VII_h$ models describe a universe rotating with an angular velocity $\omega$, and a three-dimensional rate of shear, specified by the antisymmetric tensor $\sigma_{ij}$ taken relative to the z axis. Throughout we assume equality of shear modes $\sigma = \sigma_{12} = \sigma_{13}$ [60]. Bianchi models contain a free parameter, $x$ identified by Collins and Hawking[58] which further defines the parameter h for type $VII_h$ models as [59]

$$x = \sqrt{\frac{h}{1 - \Omega_{total}}}.$$  (7.1)

where the total energy density is the sum of matter density ($\Omega_m$) and dark energy ($\Omega_\Lambda$). $x$ is related to the characteristic wavelength over which the principle axis of shear and rotation change orientation. It affects the tightness of the spiral-type temperature contribution to the CMB. Other defining parameters are the shear ($\sigma/H)_0$, vorticity ($\omega/H)_0$ and handedness $\kappa$ (-1 for left handedness and +1 right handedness).

The dimensionless shear $(\sigma_{ij}/H)_0$ is thought to be related to the dimensionless vorticity, $(\omega/H)_0$ as [59] :

$$\left(\frac{w}{H}\right)_0 = \frac{(1+h)^{1/2}(1+9h)^{1/2}}{6x^2\Omega_{tot}}\sqrt{\left(\frac{\sigma_{12}}{H}\right)_0^2 + \left(\frac{\sigma_{13}}{H}\right)_0^2}, \qquad (7.2)$$

where the different shear modes have been assumed to be equal. The dimensional shear or vorticity characterize the amplitude of the induced temperature fluctuations but not its morphology. Apart from there being both left handed and right handed models, being anisotropic, the orientation of the Bianchi pattern is flexible on the sky. Thus it allows us three additional degrees of freedom to define the model with, $(\alpha, \beta, \gamma)$ which becomes $(0°, 0°, 0°)$ when the spiral pattern corresponding to Bianchi $VII_h$ model is centered on the south pole. Hence the Bianchi $VII_h$ models can be described by a seven parameter vector $(\Omega_m, \Omega_\Lambda, x, (\omega/H)_0, \alpha. \beta, \gamma)$ and $\kappa$ separates the left handed models from the right handed ones.

The Planck team [11] simulated Bianchi CMB temperature maps using the BIANCHI2 code used by McEven et.al. [56] to simulate Bianchi $VII_h$ models. They simulated the Bianchi temperature maps for a varying $x$ and density parameters and the effect can be seen in Figure 7.4.

For the most physically motivated open-coupled-Bianchi Model where the Bianchi $VII_h$ model is coupled to standard cosmology, there is no evidence in support of a Bianchi Contribution. The Planck team however does not completely rule out Bianchi $VII_h$ cosmologies in favor of $\Lambda$CDM cosmologies.

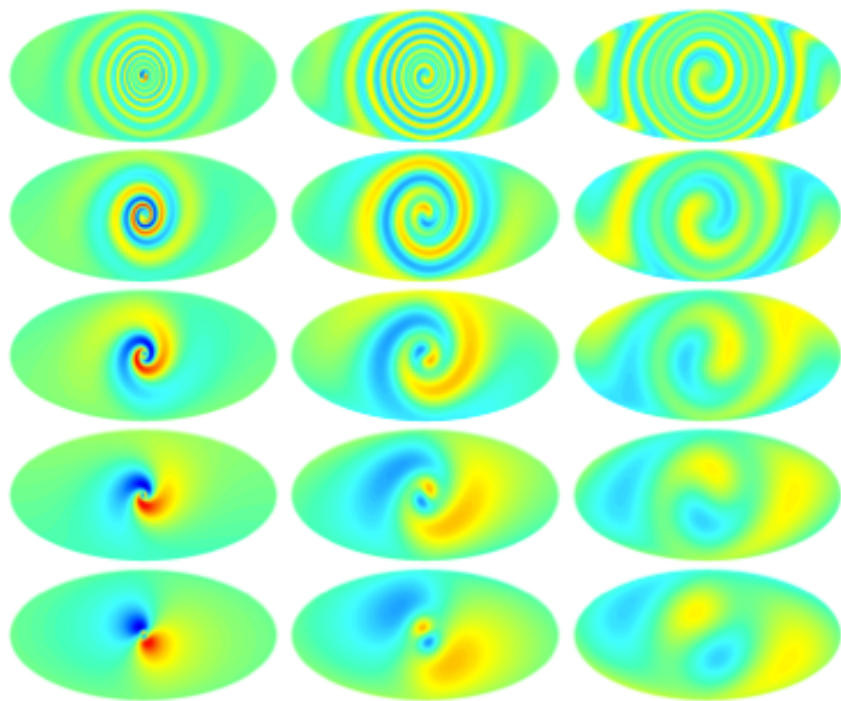

Figure 7.4: Bianchi universes with $x \in [0.1, 0.3, 0.7, 1.5, 6.0]$ from top to bottom and $\Omega_{tot} \in [0.10, 0.30, 0.95]$ from left to right [11].

# Chapter 8

# Analytical Methods to detect Multi-connectedness

A multi-connected universe consists of points which can be connected with more than one geodesic. Thus, an important observable effect of such a universe is that light will have travelled across the fundamental domain more than once if its size is lesser than the surface of last scattering. For a small enough universe, we should be able to see multiple images of a cosmic object in different directions in the sky. Looking for these multiple images is one of the most straightforward and direct methods of looking for evidence of multi-connected topology.

The possibility to beyond the fundamental cell of a multi-connected universe depends on the odds of it being comparable to the size of the observable universe. The chances are maximized when the search is conducted on the farthest possible surface that we have access to data from. The cosmic microwave background offers the best possible chances with current methods to detect a topological signature which would skew the otherwise isotropic nature of the temperature perturbations. In other words, the temperature perturbations in direction $\hat{n}$, T($\hat{n}$) would be correlated to the perturbations in another direction $\hat{m}$ by an amount that would clash with the one expected from a usual isotropic correlation function C($\theta$) defined in section 8.1, $\theta$ being the angle between $\hat{n}$ and $\hat{m}$. In a pixelized map, this results in a correlation index than does not solely depend on the angular separation between pixels. Such correlations are an important clue to detecting possible manifestations of topology.

If we live in a multi-connected universe, with the fundamental cell size smaller than the observable size of the universe, we should in principle be able to detect multiple images of the most distant cosmic objects. However, a direct search for these multiple images brings us face to face with many observational problems

like:

(i) Morphological effects when viewing from different directions,

(ii) Differentiating or finding correlations between objects at different distances which would correspond to different stages in the life of the object and

(iii) Foreground contaminations or high obscuration regions masking the object or its images.

These problems, however, can be overcome by using statistical methods, which provide a more accurate measure of the possible matches by using indicators or signatures to search for a given topology.

Another way to look for non-trivial topology is to use data from the cosmic microwave background which gives us information about the pressure density fluctuation map of the universe at the surface of last scatter. Many methods have been suggested to look for topology using the CMB and can be looked up in detail from [66]. The most popular from among these is a method that looks for matching circles in the background, called the circles in the sky method, devised by Cornish and Spergel in 1998 [55] and is further discussed in Section 8.3

## 8.1 Correlation Matrices

CMB temperature correlation matrix is a double radial integral of the average of the product of source functions that represent the transfer of the photons through the universe from the last scattering surface to the observer:

$$C_{pp'} = \int_0^{\chi_{lss}} d\chi \int_0^{\chi_{lss}} d\chi' \langle S(\chi \hat{q}_p) S(\chi' \hat{q}_{p'}) \rangle, \qquad (8.1)$$

where $\hat{q}_p$ and $\hat{q}_{p'}$ are unit vectors in the direction of the pixels p and p' while $\chi$ and $\chi'$ are proper distances along radial rays towards the last scattering surface.

There are currently two ways of computing CMB correlation functions for multi-connected universes. One of the ways to compute the CMB correlation functions for multi-connected universes would be to construct an orthonormal set of basis functions that satisfy the boundary conditions of the fundamental domain of the manifold. One then computes the spatial correlation function $\langle S(\chi \hat{q}_p) S(\chi' \hat{q}_{p'}) \rangle$ from this basis [55; 64].

The usefulness of these methods relies on the fact that in case of compactification and/or global anisotropy, the correlation function would not solely be a function of the angular separation between $p$ and $p'$ . In harmonic space, the two point correlation function is given by

$$C_{ll'}^{mm'} = \langle a_{lm} a_{l'm'}^* \rangle \neq C_l \delta_{mm'} \delta_{ll'},$$ (8.2)

where $a_{lm}$ are the spherical harmonic coefficients of the temperature function on the CMB given by

$$T(\hat{q}) = \sum_{lm} a_{lm} Y_{lm}(\hat{q}).$$ (8.3)

The Planck collaboration [11] calculated these correlation functions for different topologies and plotted them to obtain the correlation patterns typical of a certain topology. Figure 8.1 shows the correlation matrices in pixel space, showing the magnitude of correlation between the pixels. Smoothing with a gaussian field gives us a microwave background map for each of the said topologies as shown in Figure 8.2

## 8.2   Cosmic Crystallography

The most evident way to look for signatures of a multi-connected topology is to look for repeated images of distant cosmic sources. This search however is not without difficulty and involves problems such as having to take into account the time evolution of the object which may look considerably different from its image at different distances in the cosmic scenario. Morphological and foreground distortion could further skew the data and make identification of images harder that it would initially appear. It seems best to approach the problem with a statistical approach by using specific statistical signatures to look for signs of a non-trivial topology. Here, we describe the pair separation method, more commonly known as the Cosmic Crystallography method as described in detail by [26].

In 1996, Lehoucq et. al. [65] proposed the Pair Separation Histogram (PSH) method to look for correlations in the sky. The method relies upon the function of the distance, $F(d) = s$ between pairs of images in a catalogue C. Then we calculate the number of pairs whose separation lie within one of the sub intervals $J_i$, from the partitions of $[0, S_{max}]$ where $S_{max} = F(2\chi_{max})$ and $\chi_{max}$ is the survey depth of the Catalogue. The separation function is usually taken to be simply the distance between the pair $s = d$ or its square $s = d^2$. A normalized plot of the separations gives us the pair separation histogram of the data.

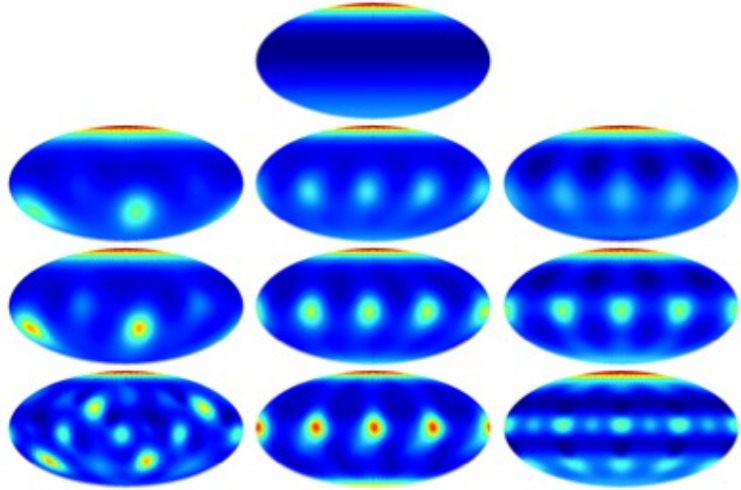

Figure 8.1: A correlation structure in pixel space of i) At the top, a simply connected isotropic universe ii) second row, from left, dodecahedral, octahedral and equal sided torii, with the fundamental domain comparable to the size of the last scattering surface iii) Subsequent rows correspond to a smaller fundamental domain with respect to the last scattering surface, from [11].

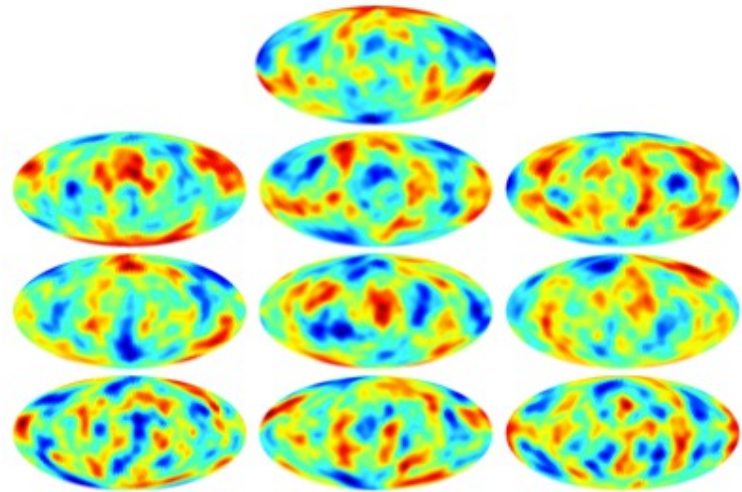

Figure 8.2: Realizations of the microwave background from the correlation maps superimposed with a gaussian profile of full width at half maximum of 640', from [11].

In more detail, for a catalogue of n cosmic sources, $\eta(s)$ denotes the number of pairs of sources whose separation is s. Divide the interval $(0, s_{max}]$ in m equal subintervals(bins) of length $\delta s = s_{max}/m$; the intervals are thus

$$J_i = (s_i - \frac{\delta s}{2}, s_i + \frac{\delta s}{2}]; i = 1, 2, ..., m \tag{8.4}$$

and centered at $s_i = (i - \frac{1}{2})\delta s$. The Pair Separation Histogram (PSH) will be given by the counting function

$$\phi(s_i) = \frac{2}{n(n-1)}\frac{1}{\delta s}\sum_{s \in J_i}\eta(s), \tag{8.5}$$

normalizable using the condition, $\sum_{i=1}^{m}\phi(s_i)\delta s = 1$.

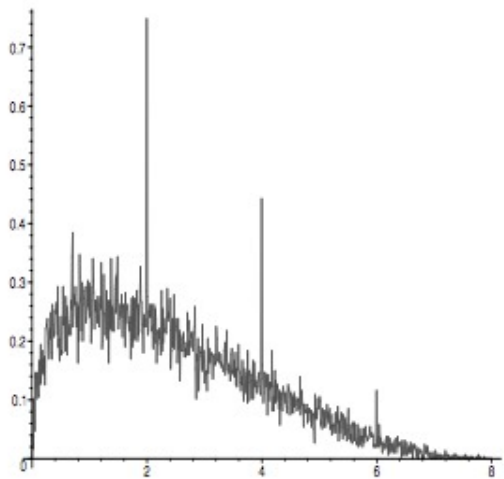

Figure 8.3: A PSH for a 3-torus topology where the horizontal axis is the separation squared while the vertical axis plots the number of pairs [26].

An example of a PSH for a universe with non-trivial topology, a 3-torus in this case, is shown in Figure 8.3. The curve characteristics are a mean curve with additional spikes which are a combination of noise and the signal. The mean curve is representative of the expected PSH, EPSH and is exclusive of statistical noise.

Further tests revealed confirmation of expected spikes in flat manifolds [65], yet no spikes were observed for hyperbolic multi-connected manifolds [67]. The reason was discovered after a theoretical statistical analysis, that the spikes of topological origin in PSH were due to only one type of isometry, the Clifford translations (CTs) which are isometries $g_t \in \Gamma$ such that for all $p \in \tilde{\mathcal{M}}$, the distance $d(p, g_t p)$ is a constant. The CTs reduce to regular translations in Euclidean spaces and the lack of CT translations in Hyperbolic geometry results in its absence of spikes in these universes.

Though other isometries also result in a slight deformation of the expected pair separation histogram as shown by [68], their amplitude is usually small as compared to the background noise and thus, they are barely detectable.

Although a good method for detecting non-trivial topology for flat manifolds and a few spherical manifolds that admit Clifford transitions, the above method fails for hyperbolic universes. Uzan et. al. [69] thereby suggested another method called the *Collecting Correlated pairs method* which uses the property of preservation of distances between images.

The method's central premise is that the distance function between two objects and their images should be equivalent, *i.e.* for each g $\in \Gamma$,

$$d(p, q) = d(gp, gq) \tag{8.6}$$

given that the pair of images(p,q) is also in the Catalogue. The analysis is basically done computationally by arranging the distances between all $P = n(n-1)/2$ possible pairs between the n sources in the catalogue and arrange them in increasing order. Then, the difference between successive distances is evaluated, with $N$ the number of times the distance equals zero or a very small number $\epsilon$ taking measurement errors into account. The CCP index, which is given by

$$\mathcal{R} = \frac{N}{P-1}, \tag{8.7}$$

is evaluated. A value of $\mathcal{R} > 0$ can be taken as a sign of multi-connectedness with higher values making the case more certain. The method however, has its own associated problems which comes from uncertainty in the determination of precise placements and distances of cosmic objects, not to forget the red shift corrections due to additional peculiar velocities of these objects. [70] discusses further limitations and uncertainties associated with the above methods. Numerous other methods are currently being developed which may provide a more accurate statistical indicator for non-triviality [71; 72; 73].

## 8.3 Matched Circles Method

The Cosmic Microwave Background is currently the most primitive source of information that we have access to and it allows us to look as far into space as possible with current technology. The information in the CMB encodes the data from redshift, $z = 1100$ which is the last scattering surface, the radius of which is given by

$$R_{sls} = R_{curv} \ arc \cosh \left( \frac{2 - \Omega_0}{\Omega_0} \right),  \tag{8.8}$$

where the photons first decoupled from matter.

The main premise of this method relies on the idea that if the observable universe around us can be thought of as a sphere of radius $\chi_{lss}$. In the scenario where the fundamental domain of our topology is smaller than this sphere, there will be an intersection between the spheres of their last scattering surfaces, and the intersection would be the shape of a circle irrespective of the curvature or structure of the topology. However, the positions and angular correlations of the circles in the sky will be determined by the topology. The matched circles will have an identical patter of temperature variations. The mappings from the surface of last scatter to the night sky is conformal thus making sure that the angles are still preserved. This implies that the identified circled appear equally resealed.

Cornish et. al. [55] developed the statistical method to detect these correlated circles in the microwave background. They selected two points, p and r in the night sky and drew circles of angular radius $\alpha$ around each point and consider all possible phases, $\phi_*$ between the points. If $T_p(\phi)$ and $T_r(\phi + \phi_*)$ are defined as the temperature fluctuations along the two circles, the circle comparison statistic is given by

$$S_{p,r}^{\pm} = \frac{\langle 2T_p(\pm\phi)T_r(\phi + \phi_*) \rangle}{\langle T_p(\phi)^2 + T_r(\phi)^2 \rangle},  \tag{8.9}$$

where $\langle \rangle = \int_0^\pi d\phi$ and the $\pm$ sign refers to whether the circles are ordered in the same or opposite directions respectively. S varies from [-1,1], being $\pm$ for perfect correlation and 0 for perfectly unmatched circles. For each radius $\alpha$, we compute the maximum possible value of the S statistic. The circles are ordered in clockwise direction for orientable and anti-clockwise for non-orientable topologies.

Better angular resolution provides a better possibility to detect these correlations as it provides a larger dataset of points. An experiment with angular resolution $\Delta\theta$ provides around $N \approx 2\pi \sin \alpha / \Delta\theta$ data points around each circle of angular radius $\alpha$.

Cornish et. al. [74] later modified the statistic such that the above computation can be realized faster by working with Fast Fourier Transforms (FFT) of the temperature correlations along the circles,

$$T_p(\phi) = \sum_m T_{p,m} exp(im\phi)$$

where $|m|$ which represents the $m$th harmonic around the $p$th circle. The statistic can be written as

$$S_{p,r}^+(\alpha, \phi^*) = \sum_m s_m exp(-im\phi_*)$$

where

$$s_m = 2 \sum_m T_{p,m} T_{r,m}^* / \sum_n (|T_{p,n}|^2 + |T_{r,n}|^2)$$

using the inverse fourier transform. The modified statistic can be written as:

$$S^+(\alpha, \phi_*) = \frac{2 \sum_m |m| T_{p,m} T_{p,m}^* e^{-im\phi_*}}{\sum_n |n| (|T_{p,n}|^2 + |T_{r,n}|^2)}, \tag{8.10}$$

which weighs the temperature coefficients with the $m$th harmonic taking into account the number of degree of freedom per mode. Doing so, enhances the contribution from the small scale structure as compared to the larger fluctuations which are dominated by the Integrated Sachs Wolf effect (ISW).

General searches explore a six parameter space: the location of the first circle centre $(\theta_p, \phi_p)$, the location of the second circle centre $(\theta_r, \phi_r)$, the angular radius of the circle $\alpha$ and the relative phase between the two circles $\phi_*$. Being excessively computationally intensive, with $N^2 \, logN$ operations, the searches performed till now, have been limited to antipodal or nearly antipodal circles. There is a large subset of topologies that could manifest in an antipodal configuration, except perhaps the Hantzsche-Wendt space for flat universe or the Picard space [75].

The searches performed by the Planck team turn up no evidence for a multi-connected topology [11]. There were no statistically significant correlations above the false detection level of $\alpha_{min} \approx 20°$. Thus any topology predicting matching pairs of back to back circles larger that $\alpha_{min}$ have been ruled out to a confidence level of 99%.

## 8.4 Bayesian Analysis

Though observational methods provide a concrete basis for ruling out or confirming a certain topological model, they are rife with measurement errors. It is thus useful to study the posterior likelihood of various models theoretically beforehand. Especially with the high resolution data from Planck released, searches for matched circles by the Planck team [11] did not provide any statistically significant match to any topology, thus making introduction of other methods mandatary.

These methods take into account that Gaussian likelihood functions apply as much to anisotropic and topological models and to the standard cosmology, albeit without isotropic correlations. The form of the likelihood function depends on the anisotropic model. For instance, for multi-connected models, the pixel-pixel correlation matrix includes anisotropic correlations while Bianchi models give a deterministic template in addition to the standard isotropic correlations for standard cosmology. It is assumed that there are no other sources of non gaussian signals.

Let us denote the model under examination by M, which could be an isotropic universe model, a Bianchi model, or a non-trivial topology. The data is denoted by vector d, containing harmonic coefficients or pixel temperatures depending on the basis functions used for representing the temperature function. Vector $\theta$ are the cosmological parameters which define M, and they can be split into two sets, $\theta_C$ which are parameters shared with the isotropic and simply connected model and $\theta_A$ which are the parameters specific to the anisotropic properties of the model. The posterior probability estimation of the parameters uses the Bayes theorem as given by

$$P(\theta/d, M) = \frac{P(\theta/M)P(d/\theta, M)}{P(d, M)},$$ (8.11)

where P($\theta/M$) is the joint prior probability of $\theta_A$ and $\theta_C$, $P(d, \theta, M) = \mathcal{L}$ is the likelihood.

The method is quite similar to the methods used in standard cosmological parameter estimation except that the above method is much more computationally expensive, given the additional complexity.

For the analysis of non-trivial topologies, the $\theta_C$ are the parameters that are shared with the concordance model while $\theta_A$ are the parameters specific to that particular topology for instance compactification lengths, the curvature parameter etc. Thus, it is usually $\theta_T$ parameters that have to be varied for different analysis. The 2-point signal correlation matrix, $C_{pp'}$ is precomputed for the topologies. Parameters such that signal amplitude, A and the orientation of the fundamental domain with respect to the sky, $\phi$ are usually maximized or marginalized.

Having defined the parameters, we define the Likelihood, which the probability that we would get a certain data vector, d along with its noise matrix, N, given a certain non-trivial topology. It is expressed as [11]

$$P(d|C[\theta_C, \theta_T, T], A, \phi) \propto \frac{1}{\sqrt{AC + N}} exp\{-\frac{1}{2}d^*(AC + N)^{-1}d\}. \qquad (8.12)$$

Usually, likelihood calculated in pixel space is not entirely accurate and all modes are not equally significant for the analysis. Thus, the quantities $d_p$, $C_{pp'}$, $N_{pp'}$ are expanded into discrete mode functions $\psi_n(p)$ orthonormalized over the pixelized sphere, tentatively with weights $w_p$ such that

$$\sum_p w(p)\psi_n(p)\psi_{n'}^*(p) = \delta_{nn'}$$

The role of the weights is in improving the accuracy of transforms on a pixelated sky. Now, the coefficients of the quantities in question can be written as:

$$d_n = \sum_p d_p \psi_n^*(p)w(p), \qquad (8.13)$$

$$C_{nn'} = \sum_p \sum_{p'} C_{pp'}\psi_n(p)\psi_{n'}^*(p')w(p)w(p'), \qquad (8.14)$$

$$N_{nn'} = \sum_p \sum_{p'} C_{pp'}\psi_n(p)\psi_{n'}^*(p')w(p)w(p'). \qquad (8.15)$$

$N_m$ modes are then selected out the series for comparison with the real signal, while other modes are marginalized. The likelihood equation now looks like this:

$$p(d|C[\theta_C, \theta_T, T], \psi, A) \propto \frac{1}{\sqrt{|AC + N|_M}} exp\{-\frac{1}{2}\sum_{n=1}^{N_m} d_n^*(AC + N)_{nn'}^{-1} d_{n'}\}. \quad (8.16)$$

The quantities C and N above only include the chosen modes $N_m$. The choice of the mode functions and their number is a trade off between the ease of inverting the C+N matrix and the useful information included in the data.

In the case of determining likelihood for Bianchi cosmologies, the analysis involves fitting a certain deterministic Bianchi template on a stochastic CMB background [56]. The parameters are denoted by $\theta_C$ for the stochastic background with a power spectrum $C_l(\theta_C)$ and $\theta_B$ for the Bianchi template. The likelihood of the model is denoted by

$$P(d|\theta_B, \theta_C) \propto \frac{1}{\sqrt{|X(\theta_C)|}} exp[-\chi^2(\theta_C, \theta_B)/2], \quad (8.17)$$

where

$$\chi^2(\theta_C, \theta_B) = [d - b(\theta_B)]^\dagger X^{-1}(\theta_C)[d - b(\theta_B)]$$

and $d = \{d_{lm}\}$, $b(\theta_B) = \{b_{lm}(\theta_B)\}$ are the spherical harmonic coefficients of the data and the template respectively, evaluated unto $l_{max} = 32$ which was determined to be enough to capture the structure of the CMB [76].

Incorporating the rotations to the Bianchi template, the likelihood is evaluated inharmonic space. The Covariance matrix $X(\theta_C)$ is determined by the mask settings and is equal to $C(\theta_C)$ without any mask applied, where $C(\theta_C)$ is the diagonal CMB covariance matrix [77]. With a partial sky mask applied however, we have

$$X(\theta_C) = C(\theta_C) + M$$

where M is a non diagonal covariance matrix for the mask. When the Bianchi template is null, the likelihood equation 8.17 reduces to the expression used for usual cosmological parameter analysis for the concordance model.

The Planck team [11] used two separate ways to do the likelihood analysis for the template, in terms of the coupling between the Bianchi template and stochastic background parameters $\theta_B$ and $\theta_C$. Though the coupled case is favored over the decoupled case by nature, it is worthwhile to examine the other case as well. For the decoupled case, they took a flat cosmological model while for an open universe was used in the coupled case for consistency with the Bianchi $VII_h$ models.

For each of the models, one can evaluate the Bayes factor to examine their credibility over the concordance model, from

$$E = P(d/M) = \int d\theta P(d/\theta, M) P(\theta/M). \qquad (8.18)$$

If the prior probability of the model is unknown, the Bayes factor can be determined by the ratio of the model's likelihoods. Jefferys [78] proposed the Jeffery's scale which employs the log Bayes factor to compare different model likelihoods. It is given by

$$\Delta \ln E = \ln(E_1/E_2)$$

and is a measure of the relative likelihood of models $E_1$ and $E_2$. A log Bayes factor $\Delta \ln E \geq 5$ is regarded as conclusive evidence of the superior significance of model $E_1$ over $E_2$ (given that $E_1 \geq E_2$). The certainity of the same reduces for a smaller Bayes factor with $\Delta \ln E < 1$ considered inconclusive.

# Chapter 9

# Conclusions

The primary motive of this thesis has been at providing a comprehensive review of the current focus of Cosmic Topology. Though the underlining principles have remained largely unchanged through the active years in the search for a defining topology of the universe, the focus of the searches has varied significantly. A crucial beginning point in topology is the discussion of the fact that Einstein's equations of general relativity only constrain local properties of space-time and do not specify in any way global properties of the universe.

One of the most evident changes from my perspective was the shift in focus from hyperbolic topologies during the time of COBE (before the discovery of dark energy), which provided evidence ($\Omega_{tot} = 0.3$) that the universe might have a negative curvature, to now, where the focus is on positively curved topologies, the Poincare Dodecahedron being the most notable, when the odds lie more towards a positively curved topology ($\Omega_{tot} = 1.02$), though not entirely ruling out a hyperbolic one. Another noticeable evolution has been in the kind of methods that have been used to investigate multi-connected topology. Having started out as a purely theoretical speculation, cosmologists gradually devised and developed observational tests as the skies were mapped to an unprecedented degree of resolution and it became possible to measure spatial curvature, spectral index etc thereby quantifying parameters for quantitative analysis of various topologies. This enabled cosmologists to put reliable constraints on cosmological parameters for different topologies and test them against observations.

The theoretical methods focussed on by the Planck team were used to quantify relative likelihood of topologies and garner theoretical evidence in favor of or ruling out different topologies. Identifying in advance the topologies which agree to a reasonable degree with parameters obtained from observation, gives us a better chance at detecting those topologies by looking for specific signatures unique to the topology. The methods used by the Planck team, be it evaluating likelihoods or searching for circles in the sky, did not succeed in providing any conclusive evidence for any particular non trivial topology.

The universal manifold is often likened to a musical instrument, which only plays certain harmonics thereby offering a possibility to know its shape and size from merely observing the sounds it produces. The strings here are analogous to the photon baryon plasma at the surface of last scatter and the sound to the acoustic oscillations present in the plasma. A straightforward inference of this is that the harmonics of the oscillations that existed in it had to be limited by the shape of the manifold they existed in. That these harmonics should still be detectable from the CMB photons that emanated from the plasma, is a reasonable assumption, given that it is not masked by foreground or perturbations collected on the way.

The possibility of the anomalies of the low quadrupole and planar octopole being manifestations of a non-trivial topology is a promising evidence of the premise that there may be detectable local signatures of the universal manifold. However until some more ingenious methods, like the brilliant circles in the sky method, are developed to determine our topological envelope measurable from earth or somewhere in its proximity, we can only rely on technological enterprise to further our investigations. Future missions to map the microwave background should be able to provide us with higher resolution data with a better signal-to-noise ratio which will further rule out some topologies and bring others into perspective.